\documentclass[%
reprint,
nofootinbib,
amsmath,amssymb,
aps,
]{revtex4-2}

\newcommand{\p}{\partial}
\usepackage{graphicx}% Include figure files
\usepackage{dcolumn}% Align table columns on decimal point
\usepackage{bm}% bold math
\usepackage{hyperref}% add hypertext capabilities
\hypersetup{colorlinks = true, linkcolor = blue, citecolor = blue, urlcolor=blue}
\usepackage{epstopdf}

\newcommand{\bmth}[1]{\mbox{\boldmath $#1$}}
\newcommand{\grad}{\bmth{\nabla}}

\begin{document}

\preprint{APS/123-QED}

\title{Interfacial dynamics and pinch-off singularities \\ for axially symmetric Darcy flow}% Force line breaks with \\

\author{Liam C. Morrow}
\affiliation{%
	School of Mathematical Sciences,
	Queensland University of Technology, Brisbane QLD 4001, Australia\\
}%

\author{Michael C. Dallaston}
\affiliation{ %
		School of Computing, Electronics and Mathematics,
		Coventry University,
		Coventry CV1 5FB, United Kingdom \\
}%

\author{Scott W. McCue}%
\email{scott.mccue@qut.edu.au}
\affiliation{%
		School of Mathematical Sciences,
		Queensland University of Technology, Brisbane QLD 4001 Australia\\
	}%

\date{\today}% It is always \today, today,

\begin{abstract}
We study a model for the evolution of an axially symmetric bubble of inviscid fluid in a homogeneous porous medium otherwise saturated with a viscous fluid. The model is a moving boundary problem that is a higher-dimensional analogue of Hele-Shaw flow. Here we are concerned with the development of pinch-off singularities characterised by a blow-up of the interface curvature and the bubble subsequently breaking up into two; these singularities do not occur in the corresponding two-dimensional Hele-Shaw problem. By applying a novel numerical scheme based on the level set method, we show that solutions to our problem can undergo pinch-off in various geometries.  A similarity analysis suggests that the minimum radius behaves as a power law in time with exponent $\alpha = 1/3$ just before and after pinch-off has occurred, regardless of the initial conditions; our numerical results support this prediction.
%This self-similar scaling is also observed in
Further, we apply our numerical scheme to simulate the time-dependent development and translation of axially symmetric Saffman-Taylor fingers and Taylor-Saffman bubbles in a cylindrical tube, highlighting key similarities and differences with the well-studied two-dimensional cases.

	\begin{description}
		\item[PACS numbers]
		47.56.$+$r, 47.15.gp, 47.55.df, 47.55.N-, 47.55.Mh
	\end{description}
\end{abstract}

\maketitle

\section{Introduction \label{sec:Introduction}}

Studies of interfacial flows in porous media are motivated by various important applications such as oil recovery and salt-water intrusion \cite{Bear2013}. There is particular interest in the physics community in studying immiscible flows that involve a fluid of lower viscosity displacing a fluid of higher viscosity since, in this scenario, the interface is unstable and prone to viscous fingering-type instabilities \cite{Homsy1987}. This paper provides a theoretical and numerical analysis of a one-phase model for flow in a homogeneous porous medium where motion of the more viscous fluid is governed by Darcy's law
\begin{equation}
\bmth{q} = -\frac{K}{\mu}\grad p,
\label{eq:darcy}
\end{equation}
while any pressure drop due to the less viscous (or inviscid) fluid's motion is ignored.  Here $\bmth{q}$ is the Darcy velocity, which can be thought of as a locally averaged velocity, while $p$ is the fluid pressure.  The two constants $K$ and $\mu$ are the porous medium's permeability and the fluid's viscosity, respectively. The velocity of the fluid is related to the Darcy velocity via
\begin{align}
\bmth{v} = \frac{\bmth{q}}{\varphi},
\end{align}
where $\varphi$ is the porosity. We shall be concerned with flows for which there is a region of inviscid fluid occupying the domain $\Omega(t)$ (which we refer to as a bubble), so that the boundary $\p \Omega(t)$ is a moving interface to be determined as part of the problem.

As part of the problem specification, an interfacial condition on the pressure is required. We apply the macroscopic surface tension condition
\begin{equation}
	p = p_B(t)-\gamma \kappa \quad\mbox{on}\quad \partial \Omega,\label{eq:dynamic}
\end{equation}
where $p_B$ is the pressure of the inviscid bubble, $\gamma$ is a dimensional surface tension and $\kappa$ is the mean curvature of the interface.  While such a condition is used extensively in the context of two-dimensional Hele-Shaw models (which we shall discuss at length below), some justification of \eqref{eq:dynamic} is required. In real porous media, capillarity acts at the pore scale; for invasion by a non-wetting fluid, this leads to fractal-like fingering patterns on the scale of the individual pores, for which a macroscale, homogenised model is inadequate. On the other hand, there is significant experimental and theoretical evidence that for an advancing wetting fluid, the system exhibits an effective surface tension effect on a scale much larger that the pore size (see \cite{Cieplak1990,Martys1991,Rangel2009} and references therein). Macroscopic surface tension \eqref{eq:dynamic} has been used in studies of interfacial porous media flow since \citet{Chuoke1959}, and more recently by many others \cite{Ambrose2012,Ambrose2013,Brandao2018,Brener1993,Ceniceros2000,Dias2013,Levine1992,Vondenhoff2009}.  We note that due to its origin, the boundary condition (\ref{eq:dynamic}) may be more appropriate when an interface is advancing, but not receding, or vice versa.  However, due to its simplicity and ubiquity in other studies, we will use it here in both cases. In the context of the fluid-fluid interface undergoing pinch-off, assuming the viscous fluid is the wetting fluid, the viscous fluid is advancing near the neck and so the inclusion of surface tension in \eqref{eq:dynamic} is reasonable.

We are particularly interested in scenarios for which the bubble $\Omega(t)$ develops a thin neck that subsequently pinches off and breaks into two parts.  This change in topology involves a singularity in curvature of $\partial\Omega$, the type of which has been studied at length both mathematically and experimentally for a range of phenomena in fluid mechanics, including the break-up of inviscid bubbles \cite{Eggers2007,Thoroddsen2007,Fontelos2011}. Typically these studies are centred around seeking a self-similar form describing the collapse of the neck thickness by the power law $h_{\min} \propto (t_0 - t)^\alpha$, where $t_0$ is the time at which pinch off occurs, and the similarity exponent $\alpha$ is a measure of the rate at which the velocity of the interface blows up. We refer the reader to \cite{Eggers1997,Eggers2008,Eggers2015} for a comprehensive overview of the analysis of these types of finite-time singularities in partial differential equations.

A further motivation for our study is to compare and contrast three-dimensional results (or at least axisymmetric results) for immiscible flow in a porous medium with the well-studied two-dimensional problem of flow through a Hele-Shaw cell.  The analogy between these two problems arises because flow of a viscous fluid through a Hele-Shaw cell is also governed by Darcy's law (\ref{eq:darcy}), except that the parameter $K$ in (\ref{eq:darcy}) is replaced by $b^2/12$, where $b$ is the small separation distance between the two parallel Hele-Shaw plates.  To make direct comparisons, we consider (axisymmetric versions of) flow geometries that have received significant interest in the Hele-Shaw case.  These include bubble contraction due to inviscid fluid being extracted from a point~\cite{Dallaston2013,Entov2011,Lee2006}, displacement of viscous fluid in a channel geometry leading to Saffman-Taylor fingers \cite{Mclean1981,Saffman1958}, propagation of finite (Taylor-Saffman) bubbles along a channel~\cite{Taylor1959,Tanveer1986}, and injection of inviscid fluid at a point leading to tip-splitting and viscous fingering patterns in a radial geometry \cite{Paterson1981,Li2009,Morrow2019a}.

An additional observation to make about (\ref{eq:dynamic}) is that our porous media flow model is also relevant for Stefan problems which describe certain melting or freezing phenomena; in that context, the boundary condition (\ref{eq:dynamic}) is appropriate (here $p$ would need to be interpreted as temperature) as it describes the Gibbs-Thomson condition that relates melting/freezing temperature to the curvature of the solid-melt interface \cite{Back2014,Font2013,Florio2016,McCue2009,Ribera2016}.

A further theoretical motivation for our work is to study the effect that the extra component of curvature has on the flow when we move from two dimensions to an axially symmetric geometry.  To take an example, we note that while Hele-Shaw bubbles can undergo pinch-off \cite{Lee2006,Entov2011}, this change in topology does not involve the same singularities in curvature as it does for three-dimensional flows (in two dimensions, the curvature of the interface does not blow up in the limit that pinch-off is approached); as such, a number of fundamental features of our examples will not be simple extensions of the Hele-Shaw cases.   From a mathematical perspective, these issues are reminiscent of the differences between curve shortening flow and mean curvature flow in differential geometry \cite{Colding2015,Topping1998}.  Given that Hele-Shaw flow may be interpreted as a nonlocal version of curve shortening flow in the plane \cite{Chen1993,Dallaston2013,Dallaston2016}, a final reason to use \eqref{eq:dynamic} is that our three-dimensional moving boundary problem can be considered a nonlocal version of the mean curvature flow.  Again, in this context we are interested in studying the effect of surface tension in \eqref{eq:dynamic} on the development of axially symmetric curvature singularities.

\begin{figure}
	\centering
	\includegraphics[width=0.465\linewidth]{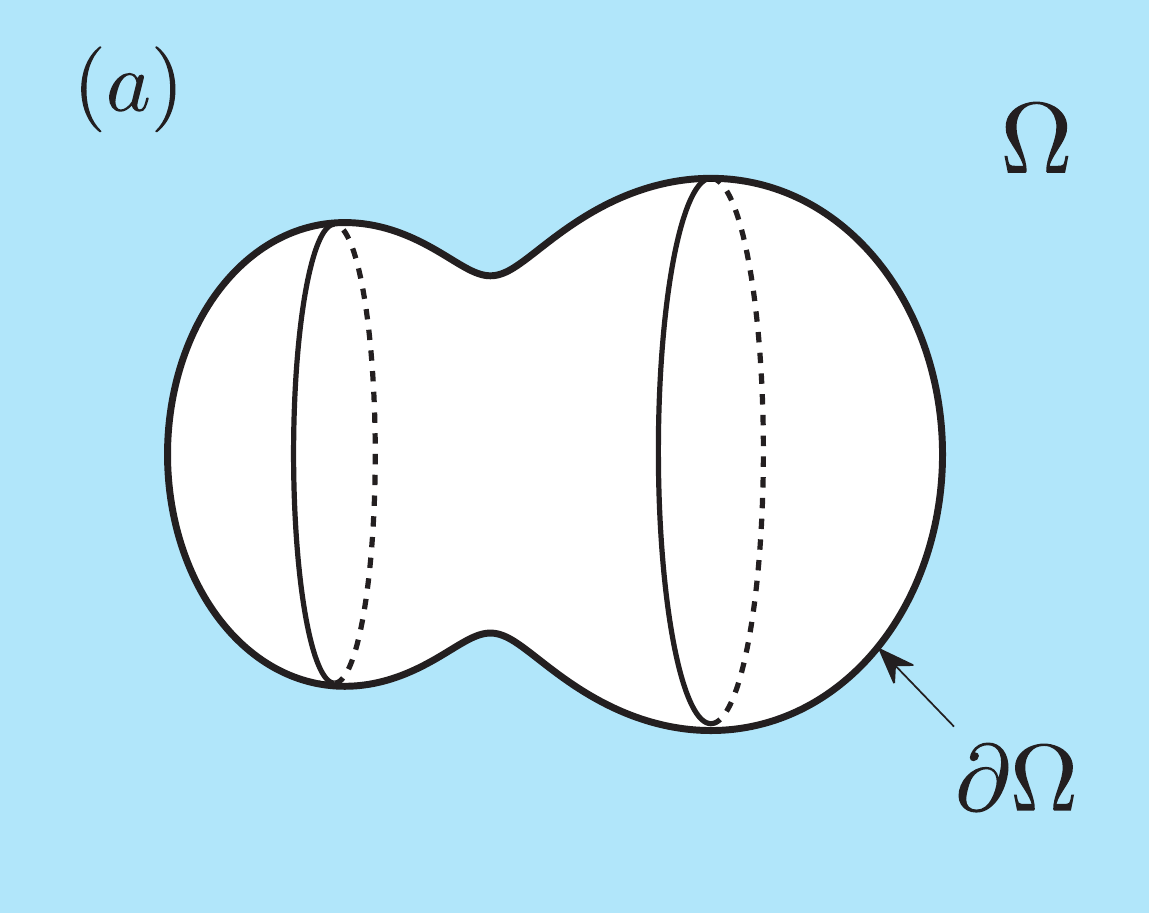}
    \hspace{2ex}\includegraphics[width=0.49\linewidth]{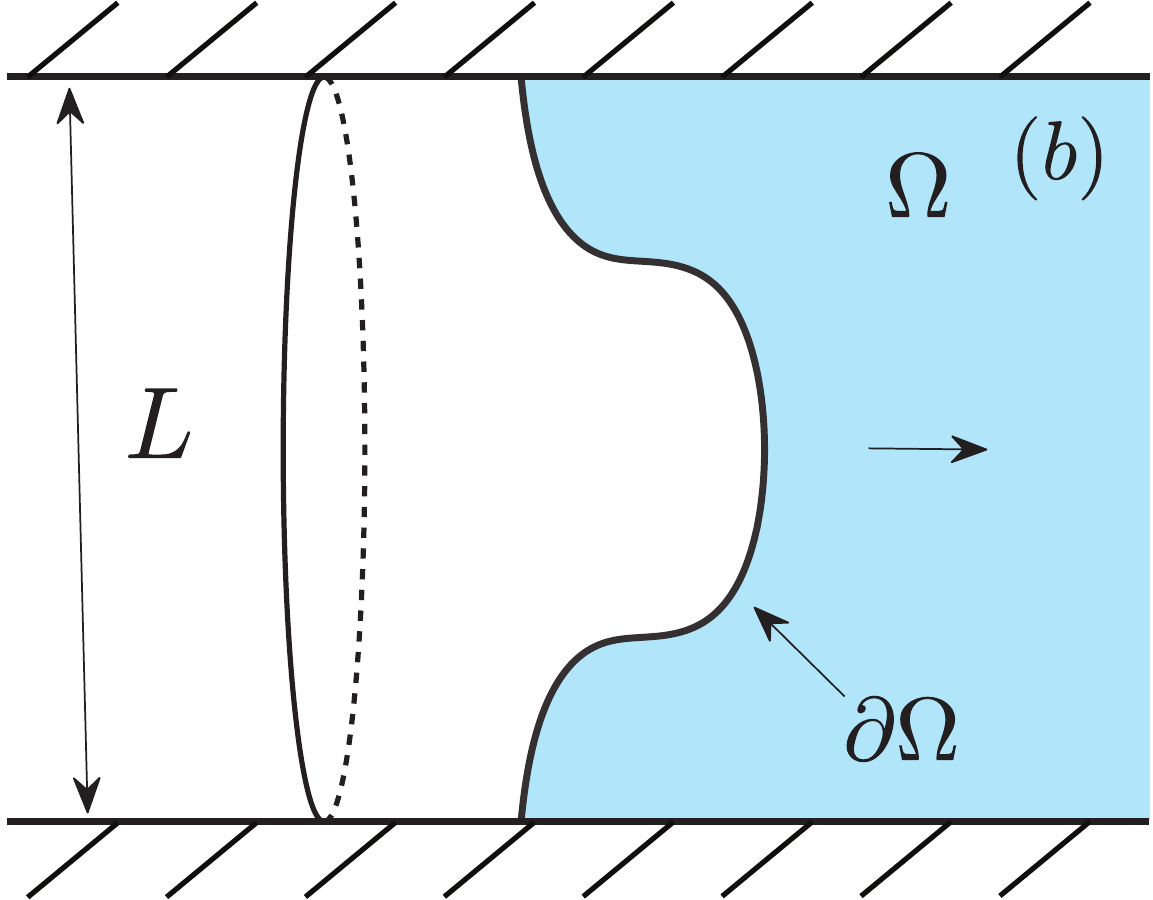}\\
    \vspace{2ex}
	\includegraphics[width=0.465\linewidth]{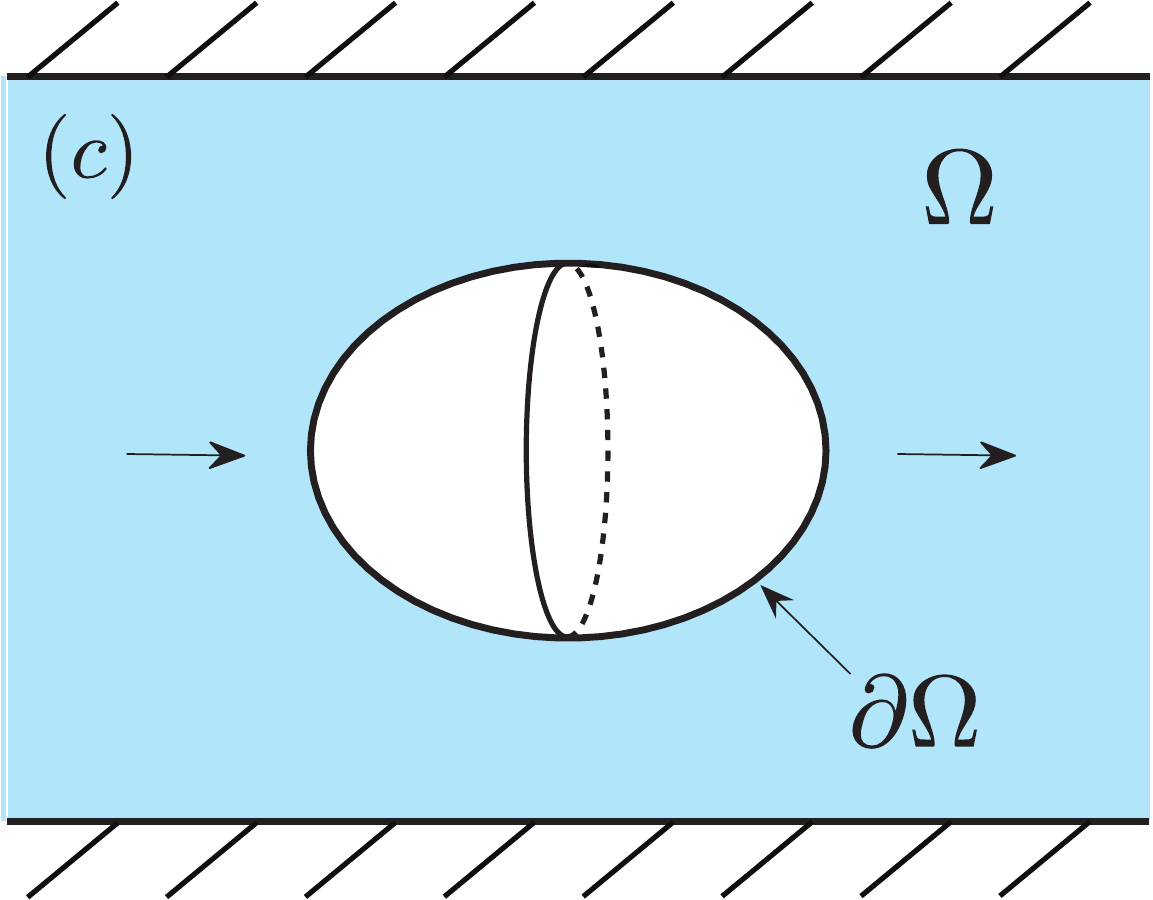}
    \hspace{2ex}\includegraphics[width=0.49\linewidth]{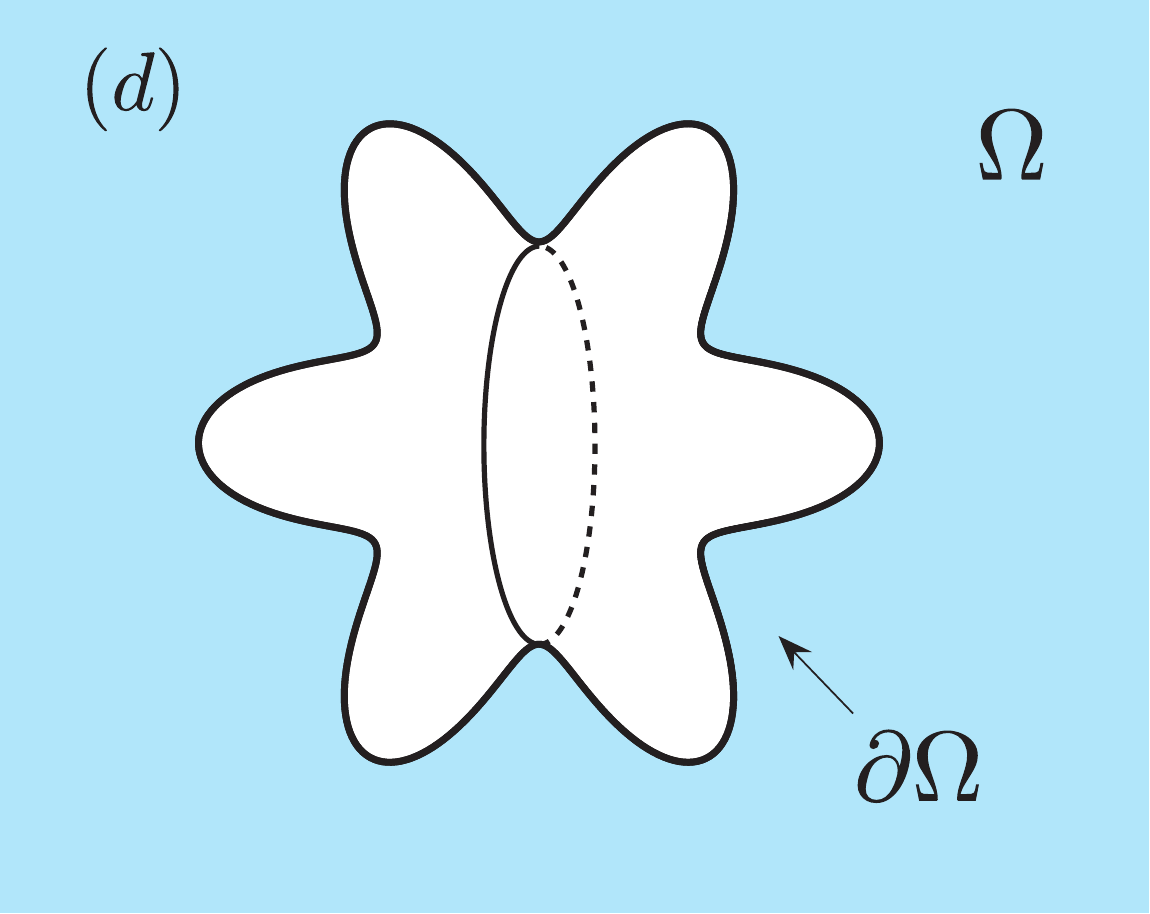}
	\caption{An illustration of the four geometries considered in this article. The blue represents the region filled a viscous fluid, and the white region denotes an inviscid bubble. $(a)$ Radial geometry, with the bubble contracting due to either the inviscid fluid extracted at a point or the viscous fluid injected at $r=\infty$. $(b)$  Cylindrical tube, with the interface propagating from left to right due to the inviscid fluid being injected at $z=-\infty$.  $(c)$ Cylindrical tube, with a finite bubble translating in the positive $z$-direction due to the viscous fluid being extracted at $z=\infty$.  $(d)$ Radial geometry, with the bubble expanding due to either the inviscid fluid injected at a point or the viscous fluid extracted at $r=\infty$.}
	\label{fig:Geometry}
\end{figure}

We begin our article in Sec.~\ref{sec:ModelFormulation} by summarising our model and explaining the connection to Stefan problems.  The four flow geometries we consider are illustrated in Fig.~\ref{fig:Geometry}.  In Sec.~\ref{sec:Radial} we treat the first of these geometries which involves a contracting bubble surrounded by an infinite body of fluid, as shown in Fig.~\ref{fig:Geometry}$(a)$.  This radial configuration is used extensively in the Hele-Shaw literature.  Our numerical scheme, based on the level set method, is summarised for this radial geometry.  The contracting bubble can eventually break up into two if the initial condition is sufficiently nonconvex.  Following the methodology of Eggers \textit{et al}.~\cite{Eggers2007} normally associated with slender-body theory, we derive a similarity solution to describe the near pinch-off behaviour, predicting a similarity exponent $\alpha=1/3$.  Our numerical solutions at times just before and after pinch-off follow the same similarity scaling and match well with the actual similarity solution in the former case.  In Sec.~\ref{sec:propagating} we treat the second and third geometries in Fig.~\ref{fig:Geometry}.  These involve flow in a cylindrical channel, a set-up that is deliberately chosen as a higher-dimensional analogue of flow in a Hele-Shaw channel.  We apply our (suitably adapted) numerical scheme and explore sufficiently nonconvex initial conditions that give rise to pinch-off singularities and convex initial conditions that evolve smoothly to travelling wave solutions.  Such travelling wave solutions are axially symmetric analogues of the two-dimensional Saffman-Taylor fingers and Taylor-Saffman bubbles and thus our time-dependent study complements that of Levine \& Tu~\cite{Levine1992} (see also Brener \cite{Brener1993}), which was for steadily moving (axially symmetric) Saffman-Taylor fingers only.  Finally, our results are summarised and discussed in Sec.~\ref{sec:discussion}, where we briefly touch on the case of an expanding bubble \cite{Brandao2018,Dias2013} (see Fig.~\ref{fig:Geometry}$(d)$) which, provided a single finger is orientated along the axis of symmetry, can also lead to the same type of pinch-off as in the other geometries just mentioned.

\section{Model Formulation \label{sec:ModelFormulation}}

\subsection{Flow through a homogeneous porous medium}

Our main physical motivation is to model the withdrawal of a bubble of inviscid fluid from a homogeneous porous medium that is otherwise saturated with viscous fluid. By assuming the viscous fluid is incompressible, $\grad \cdot \bmth{q} = 0$, then from \eqref{eq:darcy} we have Laplace's equation $\nabla^2 p = 0$ in the flow domain. We scale space, time, the Darcy velocity, the velocity of the viscous fluid, and pressure with
\begin{equation}
\begin{split}
	\bmth{x} = L \hat{\bmth{x}}, \quad t = \frac{L^3 \varphi}{Q} \hat{t}, \quad \bmth{q} = \frac{Q}{L^2} \hat{\bmth{q}}, \\ \bmth{v} = \frac{Q}{\varphi L^2} \hat{\bmth{v}}, \quad p = p_B - \frac{\mu Q}{K L} \phi,
\end{split}
\end{equation}
where $\phi$ is the velocity potential satisfying $\bmth{q} = \grad \phi$, $L$ is a representative length scale (related to the initial bubble shape or, for a more general configuration, the geometry of the flow) and $Q$ is the flow rate at which the inviscid fluid is injected/withdrawn. Retaining our original variable names, the resulting one-phase model for a bubble in a porous media is
\begin{subequations}
\begin{alignat}{3}
\nabla^2 \phi &= 0, \qquad \qquad & \textrm{in } & \mathbb{R}^3 \backslash \Omega,     \label{eq:Porous1} \\
v_n &= \frac{\p \phi}{\p n} &\textrm{on }& \p \Omega,                                    \label{eq:Porous2} \\
\phi   &= \sigma \kappa &\textrm{on }& \p \Omega,                                        \label{eq:Porous3} \\
\frac{\p \phi}{\p r} &\sim \pm \frac{1}{r^2} &\textrm{as }& r \to \infty,                \label{eq:Porous4}
\end{alignat}	
\end{subequations}
where $\sigma = \gamma K / (\mu Q)$ is the non-dimensional surface tension parameter and the positive and negative signs in \eqref{eq:Porous4} correspond to the injection and withdrawal of the inviscid bubble, respectively. The boundary conditions on the bubble boundary $\partial \Omega(t)$ \eqref{eq:Porous2} and \eqref{eq:Porous3} are often referred to as kinematic and dynamic conditions, respectively. The kinematic condition \eqref{eq:Porous2} simply states that the normal velocity of the boundary is the normal velocity of the fluid. The dynamic condition \eqref{eq:Porous3} relates the pressure to the surface tension $\sigma$ via the mean curvature $\kappa$.  This equation, the dimensionless version of (\ref{eq:dynamic}), is discussed in the Introduction.  From a geometric perspective, the surface tension acts to smooth out regions of high curvature on the interface.  For the above mathematical model to make sense in the context of contracting bubbles, the inviscid fluid must be extracted from the point to which the interface ultimately evolves (or from multiple points in the case in which the initial bubble breaks up into two or more satellite bubbles).

For the special case of zero surface tension, $\sigma=0$, there has been a number of studies on contracting bubbles, focussing on the shape of the interface in the limit that it contracts to a point \cite{Di1986,Howison1986,McCue2003,Morrow2019b}, or whether the bubble breaks up into two \cite{Morrow2019b} (in two dimensions, see \cite{Lee2006,Entov1991,Entov2011,McCue2011}).  However, there has been relatively little theoretical study concerned with solutions to \eqref{eq:Porous1}-\eqref{eq:Porous4} that undergo changes in topology when the effects of surface tension are included.  For expanding bubbles, with the positive sign in \eqref{eq:Porous4}, or for the problem involving displacement of viscous fluid in a cylindrical tube, details of linear, weakly nonlinear and travelling wave analysis has been performed \cite{Levine1992,Dias2013,Brandao2018}.  However, for these geometries, no fully nonlinear time-dependent solutions to \eqref{eq:Porous1}-\eqref{eq:Porous4} (with appropriate changes due to relevant configuration) have been previously reported.

\subsection{Application to Stefan problems \label{sec:StefanProblem}}

The other physical motivation for \eqref{eq:Porous1}-\eqref{eq:Porous4} is to model the melting/freezing of a crystal dendrite, where now $\p \Omega$ represents a solid-melt interface. Typically, melting/freezing problems are modelled mathematically as a Stefan problem, where $\phi$ represents temperature, and the field equation is the linear heat equation
\begin{align} \label{eq:HeatEqn}
\frac{1}{\beta}\frac{\p \phi}{\p t} = \nabla^2 \phi     \qquad \qquad \text{in } \mathbb{R}^3 \backslash \Omega(t),
\end{align}
where $\beta$ is the Stefan number, which is the ratio of latent heat to specific heat. In the regime for which the latent heat is significantly larger than the specific heat, it follows that $\beta \gg 1 $ so that \eqref{eq:HeatEqn} reduces to \eqref{eq:Porous1} \cite{mccue3}. In this context, the far-field condition \eqref{eq:Porous4} represents the heat flux being prescribed at infinity.  Another, perhaps more suitable, boundary condition in the context of Stefan problems that the temperature in the far field is prescribed leading to
\begin{align} \label{eq:StefanFarfield}
\phi \sim \phi_{\infty} \qquad \qquad \textrm{as } r \to \infty
\end{align}
\cite{Morrow2019b}. As we demonstrate a number of times throughout the paper, the dynamics of the interface near pinch-off are dependent only upon \eqref{eq:Porous1}-\eqref{eq:Porous3}, and therefore independent of the choice of far-field boundary condition.

\begin{figure}
	\centering
	\includegraphics[width=1.0\linewidth]{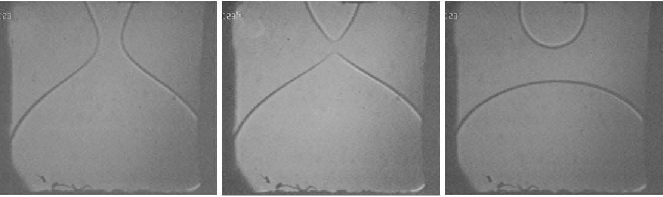}
	\caption{The time evolution of a dripping $^3$He crystal surrounded by liquid $^4$He from \citet{Ishiguro2007} reproduced with permission from the American Physical Society. The $^3$He crystal was grown in the $^4$He bath, and when it becomes sufficiently large, it is dragged down due to gravity and undergoes pinch-off. The width of each frame is 3.5 mm.  Analysis of the video footage both just before and after pinch off occurs indicates a scaling exponent of 1/3.}
	\label{fig:Figure1}
\end{figure}

Pinch-off of crystals undergoing phase change is relevant, for example, for experiments conducted by Ishiguro et al.~\cite{Ishiguro2007}. In this experiment, $^3$He crystals were grown in a $^4$He bath, and the authors argue the effects of gravity and surface tension cause the $^3$He crystal to pinch-off, as illustrated in Fig.~\ref{fig:Figure1}. It was shown that the neck of the crystal both moves towards and recoils from pinch-off with a similarity exponent of $\alpha = 1/3$.  We return to this point in Sec.~\ref{sec:discussion}.

\section{Contracting bubble geometry \label{sec:Radial}}

%\subsection{Governing equations}

\subsection{Numerical scheme} \label{sec:NumericalScheme}

In this section, we are concerned with the behaviour of a contracting bubble that undergoes pinch-off. Restricting ourselves to axially symmetric geometries in spherical co-ordinates such that $\phi = \phi(r, \theta, t)$ and with the bubble interface $\partial \Omega$ denoted by $r = s(\theta, t)$, our model \eqref{eq:Porous1}-\eqref{eq:Porous4} is
\begin{subequations}
\begin{alignat}{3}
&\frac{1}{r^2} \frac{\p}{\p r} \left( r^2 \frac{\p \phi}{\p r} \right) + \frac{1}{r^2 \sin \theta} \frac{\p}{\p \theta} \left( \sin \theta \frac{\p \phi}{\p \theta} \right) = 0 \qquad &\textrm{in } & r > s, \label{eq:Spherical1} \\
&s_t = \phi_r - \frac{1}{s^2} \phi_\theta s_\theta &\textrm{on } & r = s, \label{eq:Spherical2}\\
&\phi = \sigma \frac{3 s s_\theta^2 - \cot \theta s_\theta^3 - s^2 \left( s_{\theta\theta} + s_\theta \cot \theta \right) + 2s^3}{s (s^2 + s_\theta^2)^{3/2}} &\textrm{on } & r = s, \label{eq:Spherical3} \\
&\frac{\partial \phi}{\partial r} \sim -\frac{1}{r^2} &\text{as } & r \to \infty. \label{eq:Spherical4}
\end{alignat}
\end{subequations}

%When the bubble is expanding, it, generally, will be unstable. The development of these instabilities have been studied using both linear and weakly nonlinear stability analysis in full spherical coordinates $(r, \theta, \phi)$ \cite{Dias2013,Brandao2018}. While an axially symmetric coordinate system would be insufficient to accurately describe the growth of these fingers, in this section we are interested pinch-off singularities that develop when the bubble is contracting, which are generally considered to be locally axially symmetric.

The simulation of interfaces undergoing a topological change can be a computationally demanding task. For irrotational, inviscid flow, perhaps the most popular choice of numerical scheme is the boundary integral method (BIM) \cite{Burton2007,Day1998,Eggers2008a,Leppinen2003}. While the BIM has shown to produce accurate simulations of interfaces approaching pinch-off, it can only be used for problems where the Green's functions can be computed. Furthermore, on its own the BIM cannot easily handle changes in topology. Success has been achieved using `front capturing methods', which are generally not as accurate but have greater flexibility \cite{Dinic2019,Garzon2009,Garzon2011}. The numerical scheme we use in this article implements the level set method \cite{Osher1988}, a numerical framework that describes the evolution of moving interfaces by representing them as the zero level set of a higher dimensional hyper-surface. Our scheme was presented in \cite{Morrow2019b}, and used to study solutions to \eqref{eq:Porous1}-\eqref{eq:Porous3} and \eqref{eq:StefanFarfield} that undergo pinch-off for the special case of zero surface tension, $\sigma=0$.   In that study it was shown that our numerical scheme is capable of accurately describing the behaviour of interfaces that contract to either one or multiple points. In this section, we give a brief description of the scheme, and refer the reader to \cite{Morrow2019b} for further details.

To implement the level set method, we construct a signed distance function, $\psi(\bmth{x},t)$, where $\psi<0$ in the bubble region and $\psi>0$ in the viscous fluid region. If our bubble interface has a normal speed $V_n$, then we wish to construct a continuous function $F$ which satisfies $F = V_n$ on the interface. Thus $\psi$, and in turn the interface, evolves according to		
\begin{align} \label{eq:LevelSetEquation}
\psi_t + F |\grad \psi| = 0.
\end{align}
In the context of \eqref{eq:Porous1}-\eqref{eq:Porous4}
\begin{align}
F = \frac{\grad \phi \cdot \grad \psi}{|\grad \psi|} \qquad \bmth{x} \in \mathbb{R}^3 \backslash \Omega.
\end{align}
The spatial derivatives in \eqref{eq:LevelSetEquation} are approximated using a second order essentially non-oscillatory scheme, and integrated in time using third order Runge-Kutta. To ensure the numerical solution accurately captures the dynamics of the interface both as it approaches and recoils from pinch-off, we choose a time-step size $\Delta t = 0.05 \times \Delta x / \max|F|$. Simulations are performed on the computational domain $0 \le r \le 1.5$ and $0 \le \theta \le \pi$ using $300 \times 630$ equally spaced nodes.

\begin{figure}[h!]
	\centering
	$(a)$ Symmetric dumbbell\\
	\includegraphics[width=0.49\linewidth]{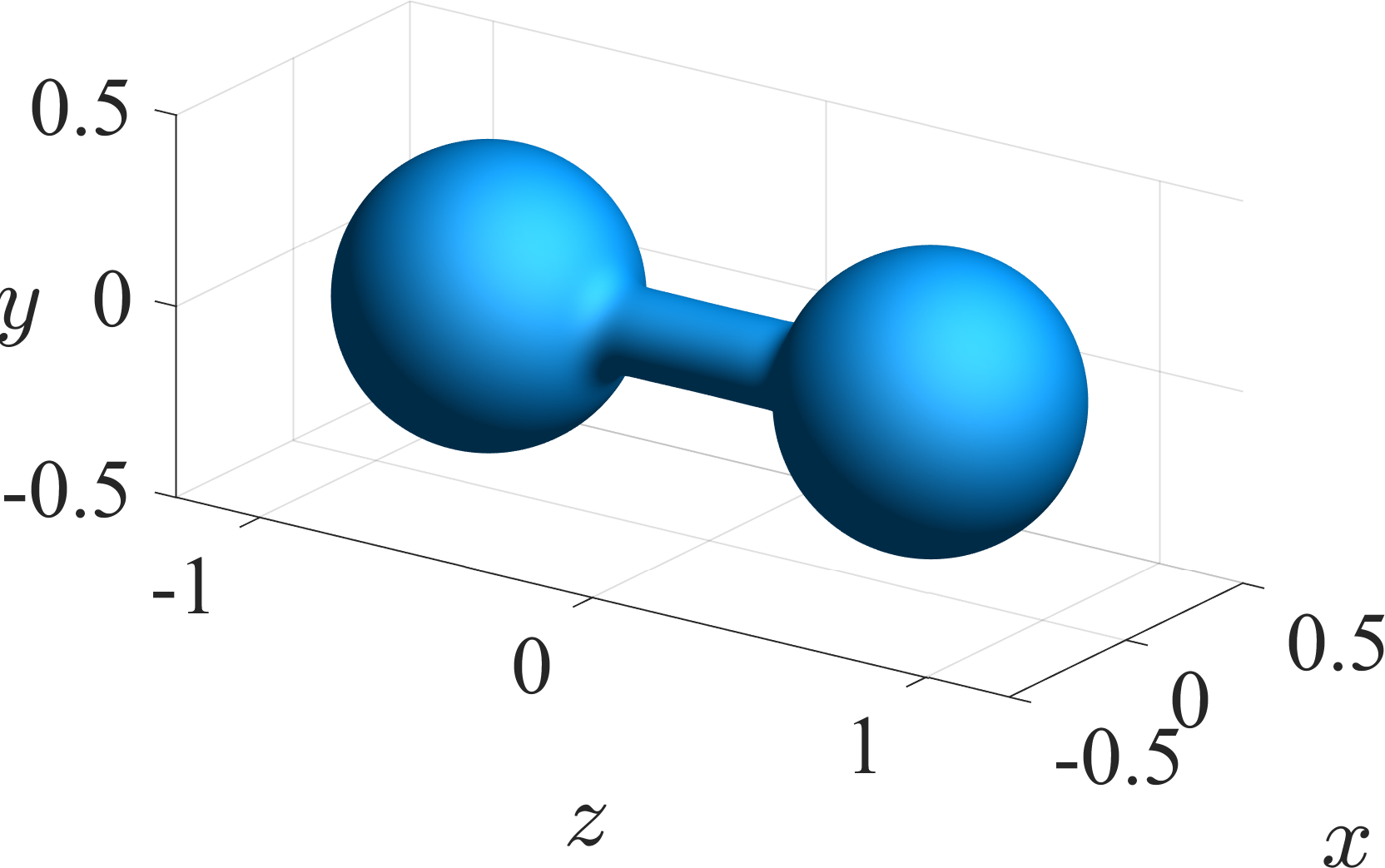}
	\includegraphics[width=0.49\linewidth]{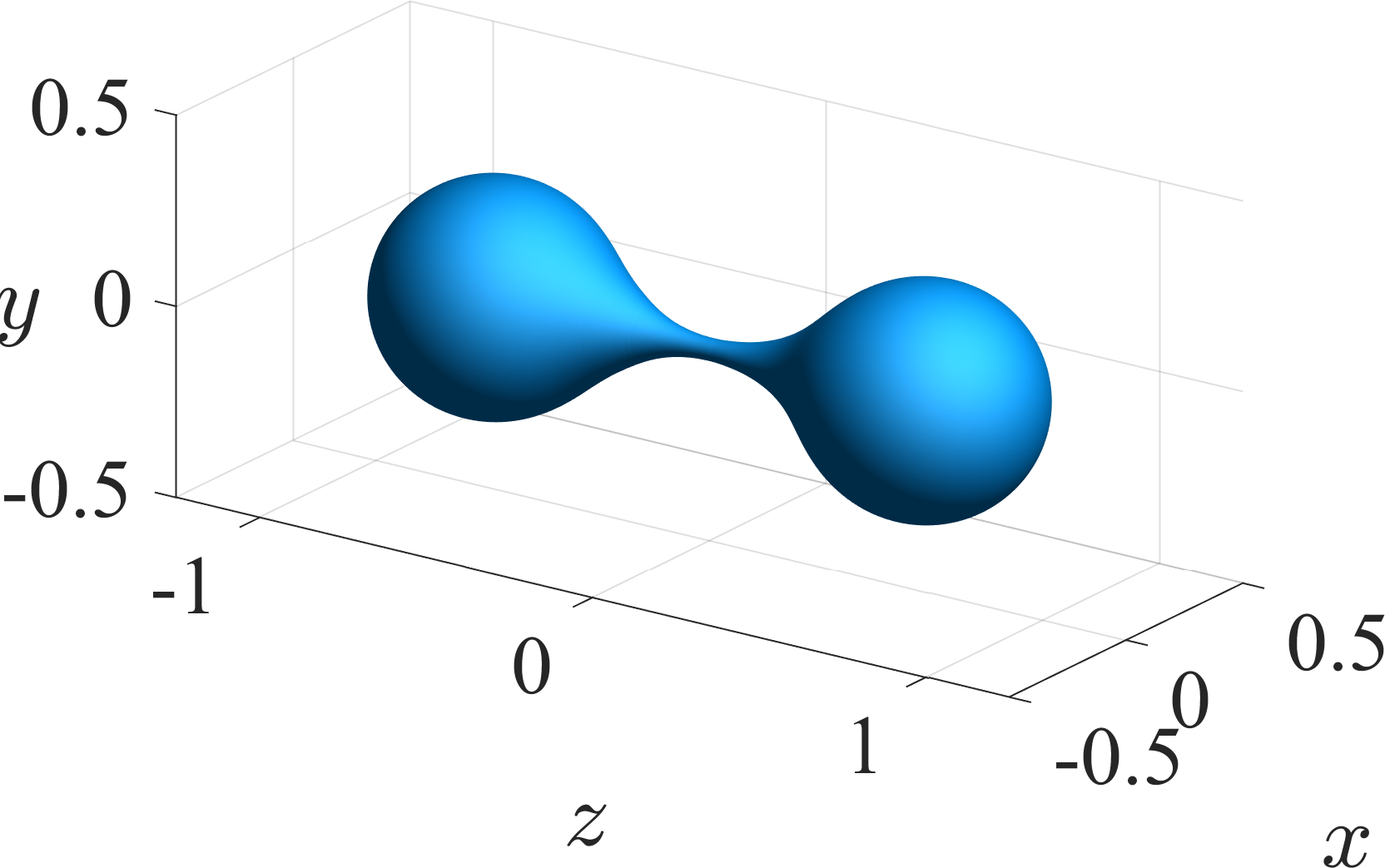}
	
	\includegraphics[width=0.49\linewidth]{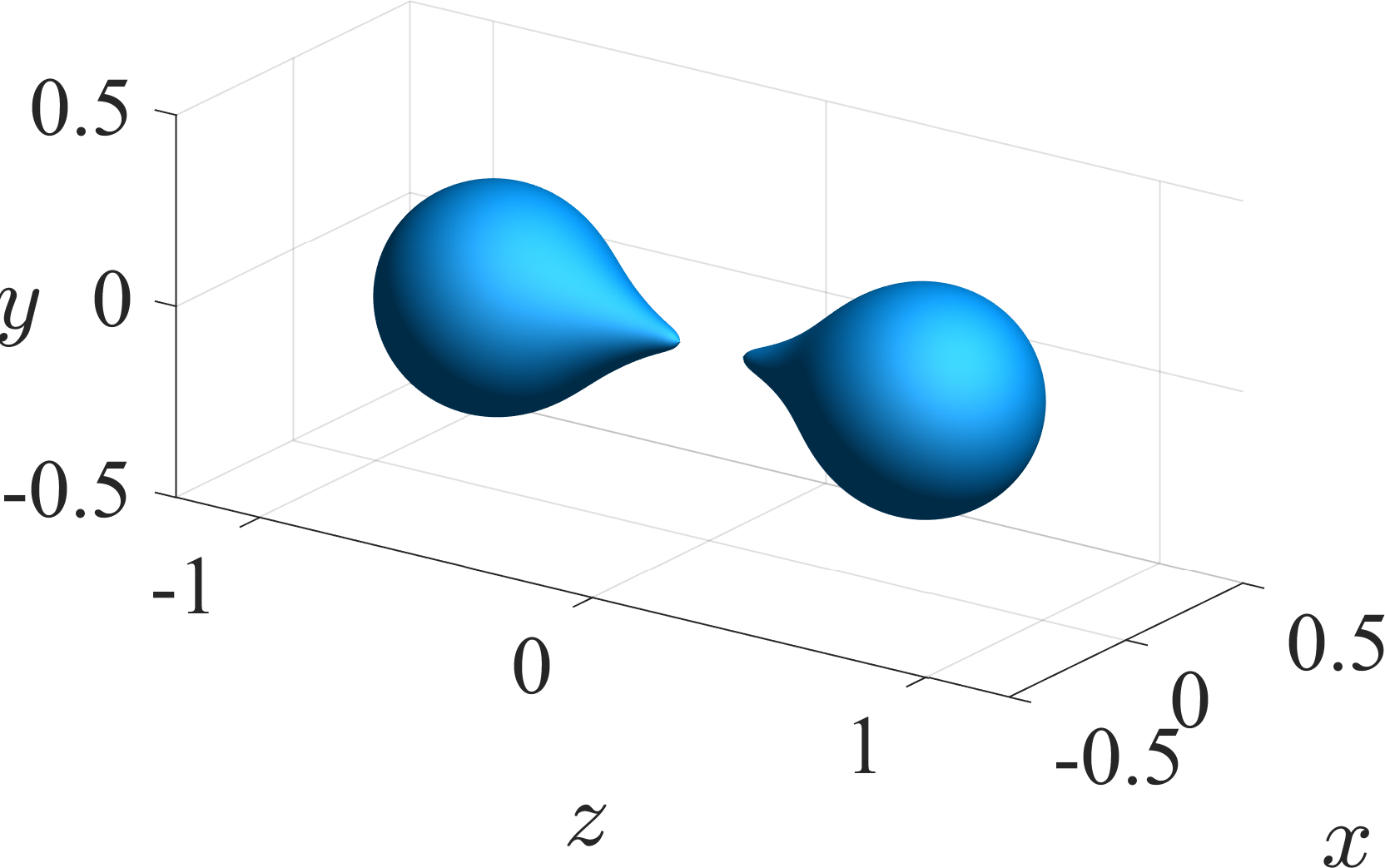}
	\includegraphics[width=0.49\linewidth]{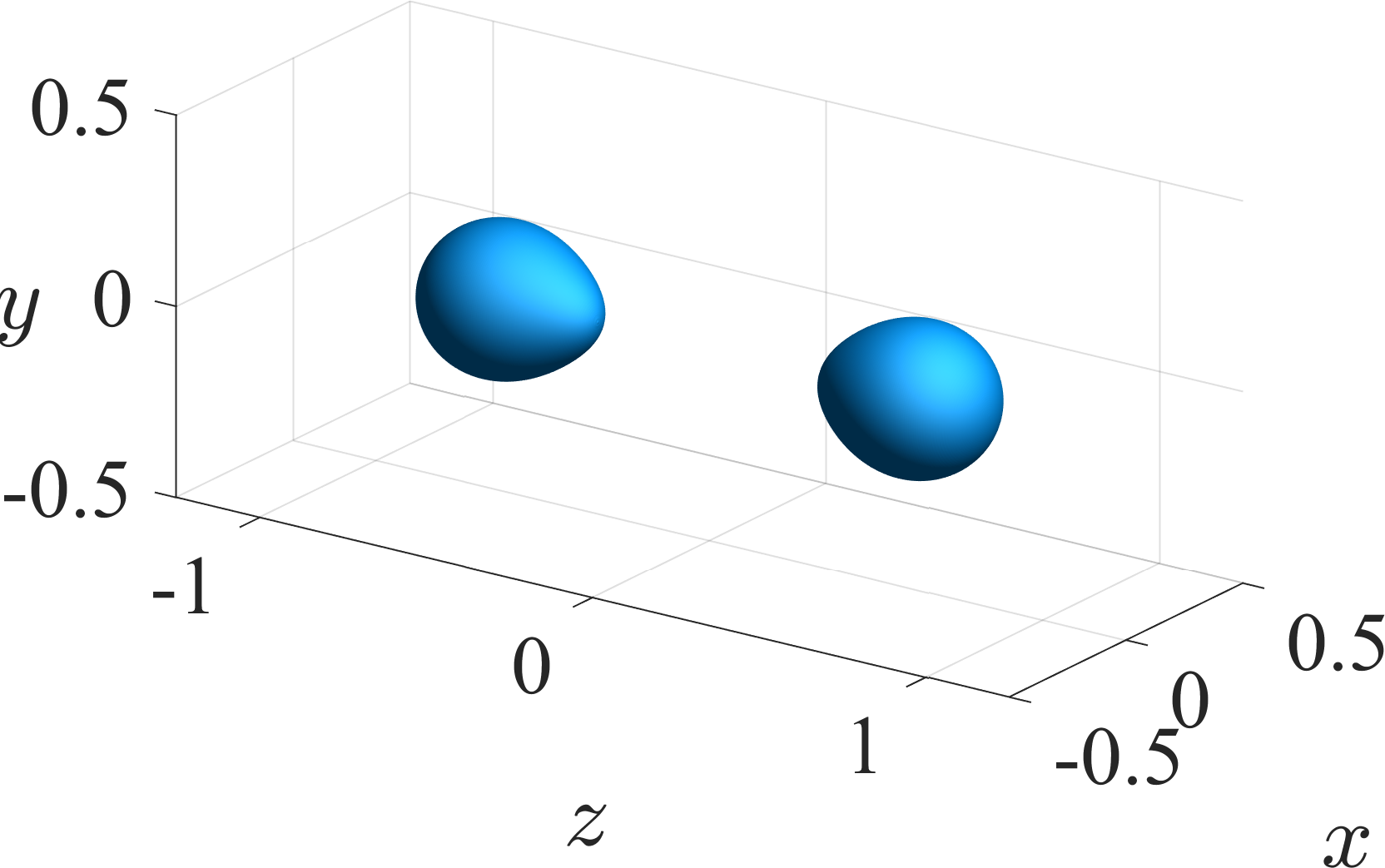}
	\\ \vspace{2em} $(b)$ Asymmetric dumbbell\\
	\includegraphics[width=0.49\linewidth]{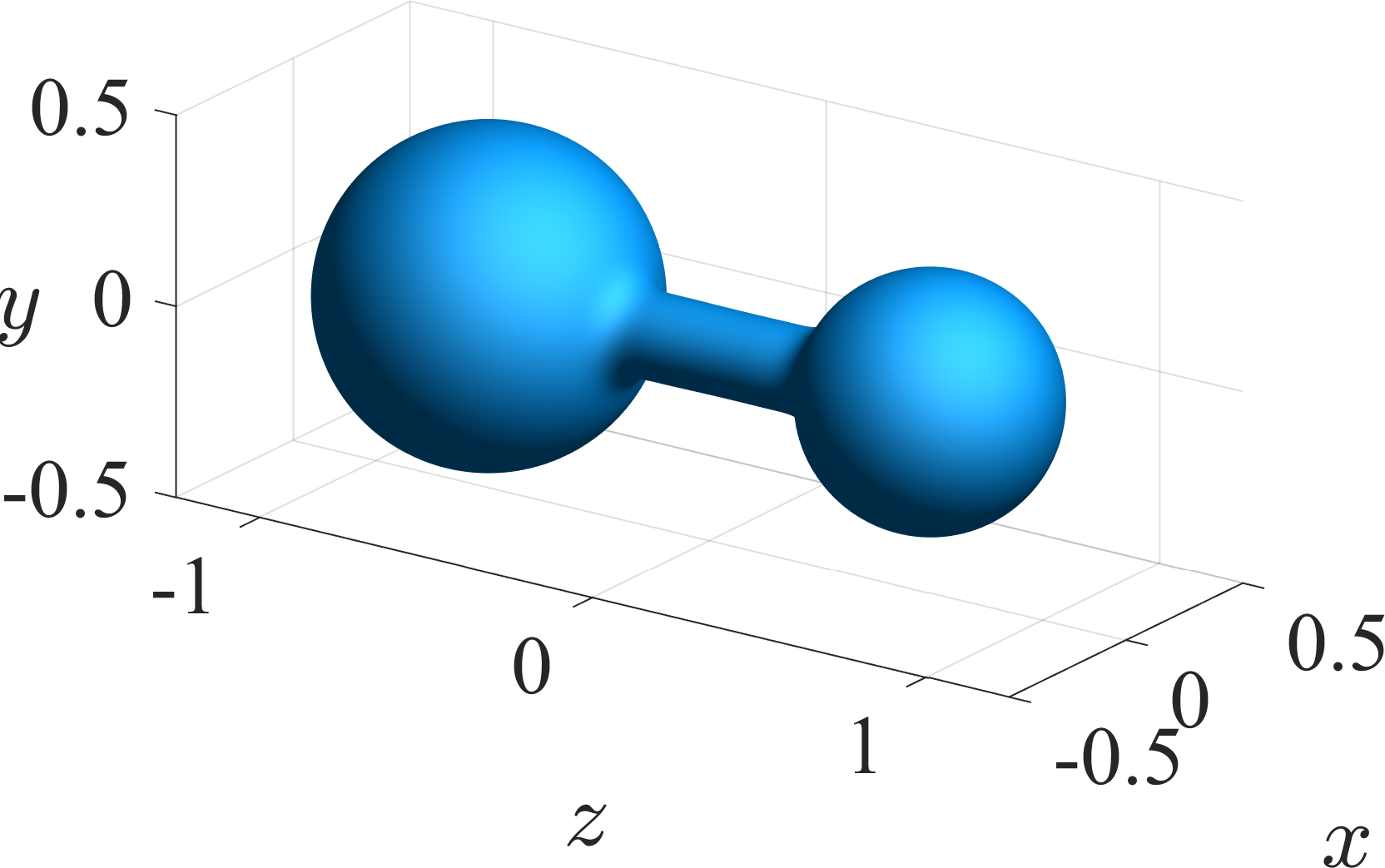}
	\includegraphics[width=0.49\linewidth]{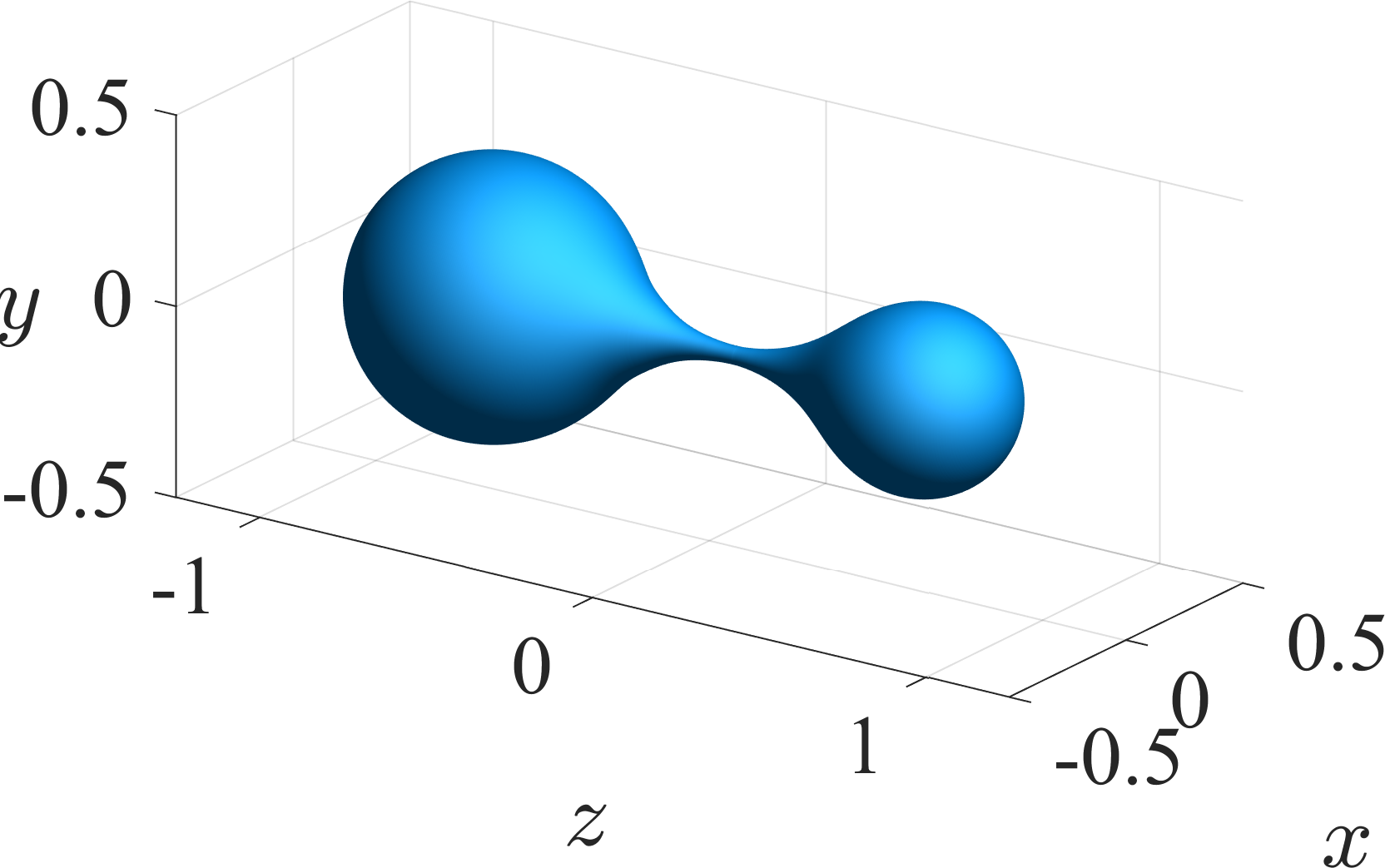}
	
	\includegraphics[width=0.49\linewidth]{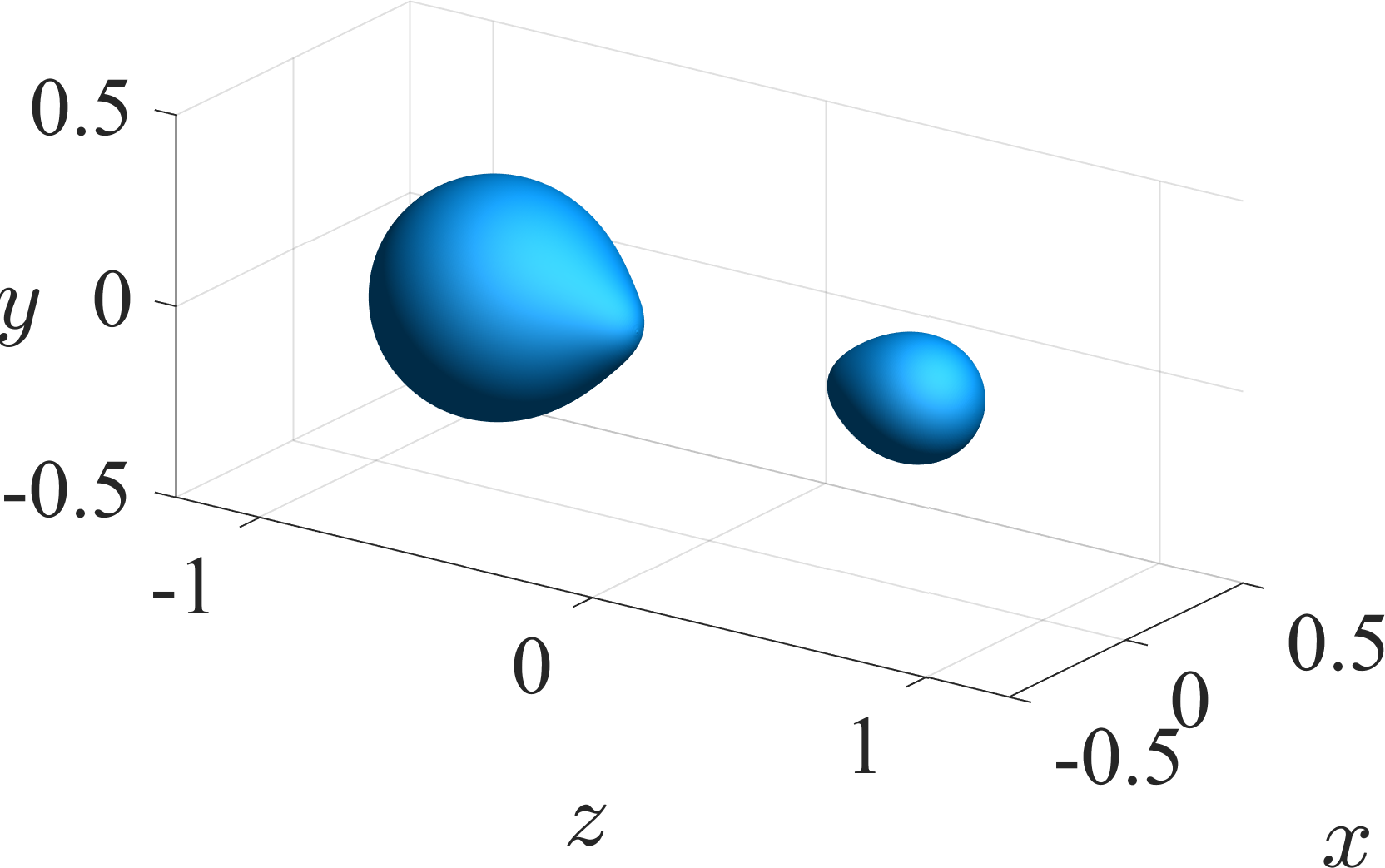}
	\includegraphics[width=0.49\linewidth]{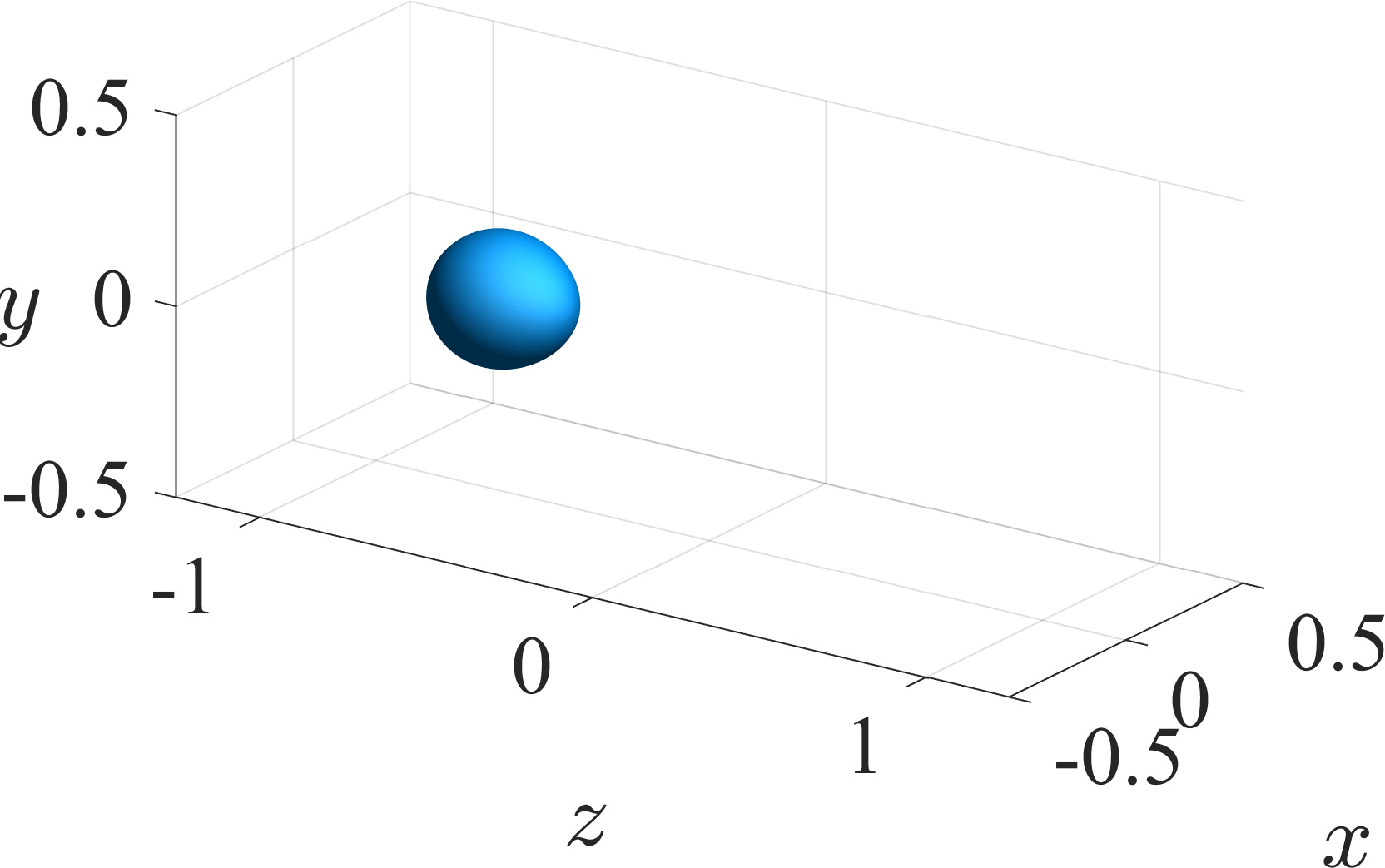}
	\caption{\label{fig:ContractingDumbbell} Numerical solution of \eqref{eq:Spherical1}-\eqref{eq:Spherical4} found using the numerical scheme described in Sec.~\ref{sec:NumericalScheme} with $(a)$ symmetric and $(b)$ asymmetric dumbbell as the initial shape of the interface. For the symmetric dumbbell, the bubble contracts and pinches-off into two satellite bubbles of equal volume that contract to a point at the same time. For the asymmetric dumbbell, when the neck of the bubble pinches off, two satellite bubbles of difference size form. The smaller of these two bubbles contracts to a point first, followed by the second. Simulations are performed on the domain $0 \le \theta \le \pi$ and $0 \le r \le 1.5$ with $628 \times 300$ equally spaced nodes.}
\end{figure}

We solve for $\phi$ using a similar procedure as described in \cite{Chen1997}. For nodes away from the interface, the derivatives in \eqref{eq:Spherical1} are evaluated using a standard five-point finite difference stencil. For nodes adjacent to the interface, the stencil is adjusted by imposing a ghost node on the interface, whose location is determined by computing points where $\psi = 0$. The value of this ghost node comes from the dynamic boundary condition \eqref{eq:Spherical3}, where the curvature term is computed using $\kappa = \grad \cdot \left(  \grad \psi / |\grad \psi| \right)$. We incorporate the far-field boundary condition \eqref{eq:Spherical4} into the stencil by implementing a Dirichlet-to-Neumann mapping \cite{Givoli2013}. While solving for $p$ allows us to compute $F$ where $\bmth{x} \in \mathbb{R}^3 \backslash \Omega$, to solve \eqref{eq:LevelSetEquation} we require an expression for $F$ over the entire computational domain. It was proposed in \cite{Moroney2017} that $F$ can be extended into $\bmth{x} \in \Omega$ by solving the biharmonic equation
\begin{align} \label{eq:Biharmonic}
\nabla^4 F = 0 \qquad \bmth{x} \in \Omega(t).
\end{align}
Computing the solution to \eqref{eq:Biharmonic} does not require the location of the interface explicitly. Instead the computational nodes that need to be included in the biharmonic stencil correspond to those where $\psi<0$. By solving \eqref{eq:Biharmonic}, this approach provides a continuous expression for $F$ over the entire domain while maintaining $V_n = F$ on the interface. We refer the reader to \cite{Moroney2017} for more details.

\begin{figure}
	\centering
	\includegraphics[width=0.80\linewidth]{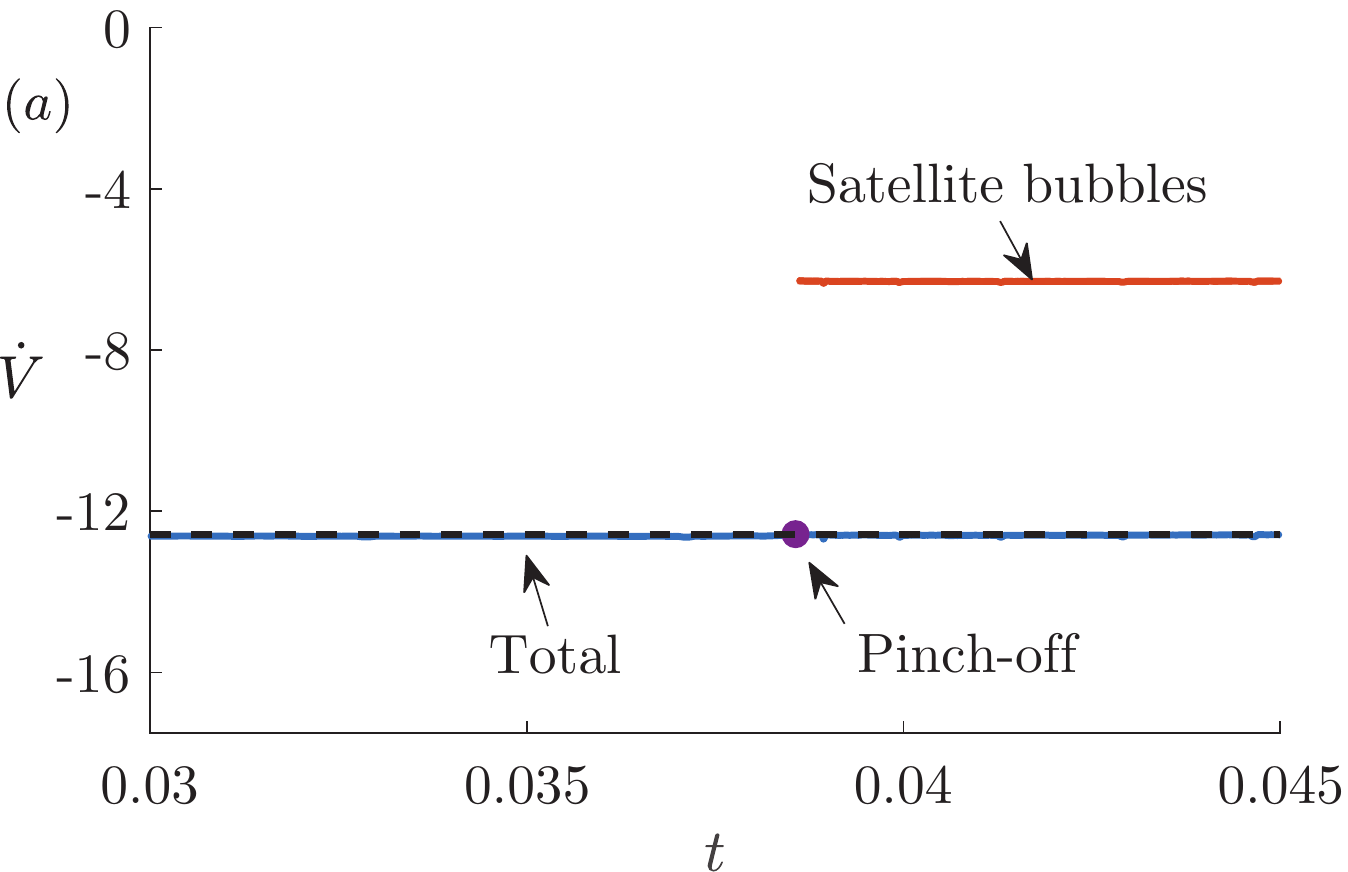} \\
	\includegraphics[width=0.80\linewidth]{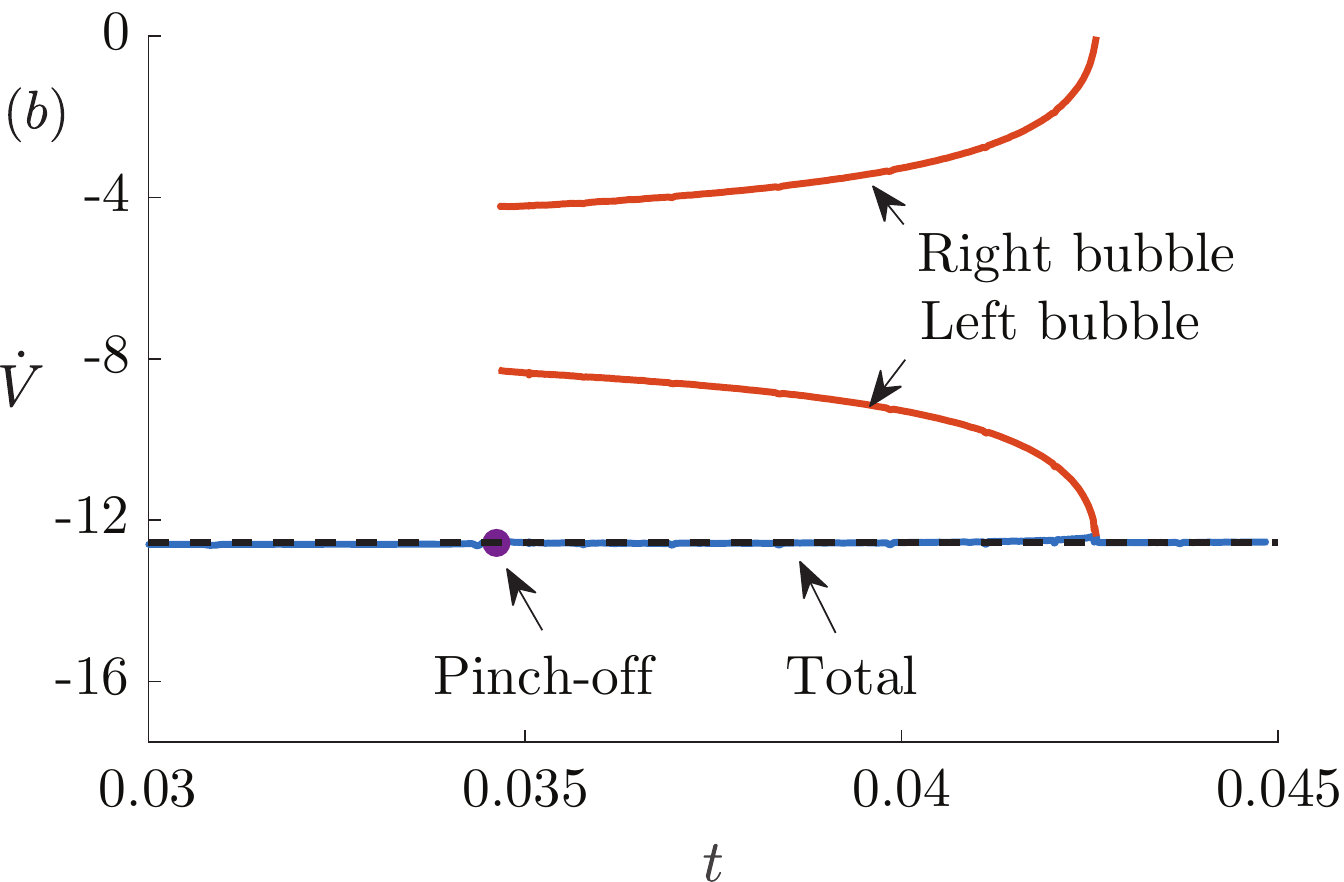}
	\caption{The rate of change of volume, $\dot{V}(t)$, of the numerical solution to \eqref{eq:Spherical1}-\eqref{eq:Spherical4}, shown in Fig.~\ref{fig:ContractingDumbbell} with $(a)$ symmetric and $(b)$ asymmetric dumbbell. The solid lines denote the rate of change of volume of the numerical solution, and the dotted (black) line denotes the exact rate of change $\dot{V} = -4 \pi$. The dots denote the point in time when pinch-off occurs).}
	\label{fig:RateOfContraction}
\end{figure}

We now provide a simple test on our numerical scheme.  As we show in Appendix~\ref{sec:ROCvolume}, the far-field boundary condition \eqref{eq:Spherical4} results in the rate of change of volume of a bubble evolving according to \eqref{eq:Spherical1}-\eqref{eq:Spherical4} to be $-4 \pi$, independently of $\sigma$. Once a change in topology has occurred, the total rate of change of volume of all bubbles will remain $-4 \pi$. To confirm this is consistent with our numerical solution, we consider the contraction rate of both a symmetric and asymmetric bubble, shown in Fig.~\ref{fig:ContractingDumbbell}, both of which undergo pinch-off. Fig.~\ref{fig:RateOfContraction} shows the rate of decrease of volume $\dot{V}$, of both satellite bubbles, the exact rate of decrease of volume, and the rate of decrease of the satellite bubbles which form post pinch-off. For the symmetric bubble, Fig.~\ref{fig:RateOfContraction}$(a)$ indicates that the total rate of change of volume of the numerical solution compares well with the expected rate. Furthermore, the rate of decrease of the satellite bubbles are both $\dot{V} = -2 \pi$, as expected. For the asymmetric bubble, Fig.~\ref{fig:RateOfContraction}$(b)$ shows that the total volume decreases at the expected (constant) rate both before and after pinch-off occurs. %Interestingly, the decrease of each satellite bubble is time-dependent, and is proportional to the volume of each bubble compared to the total volume\footnote{Liam: This statement suggests to me you can solve for each volume analytically.  Is that right?  For example, do you get $\dot{V}\propto (V(0)-4\pi t)^{-(1+1/4\pi)}$.}.

We note that our model assumes the pressure of each bubble remains equalised between the two disconnected interfaces. In turn, this can lead to counter-intuitive volume changes. For example, if the inviscid bubble is neither injected or withdrawn ($Q=0$) such that $\dot{V} = 0$, and if the bubble pinches-off into two satellite bubbles of unequal volume, as shown in Fig \ref{fig:ContractingDumbbell}$(b)$ for example, then it is expected that the smaller bubble will contract to a point. In turn, this would result in the larger bubble expanding. However, as our primary motivation is to investigate the behaviour of solutions to \eqref{eq:Porous1}-\eqref{eq:Porous4} both approaching and recoiling from a pinch-off singularity, we expect that our model is still reasonable.

\subsection{Pinch-off for a contracting bubble} \label{sec:Contracting}

\subsubsection{Similarity exponents}\label{sec:similaritysoln}

We wish to study solutions like those presented in Fig.~\ref{fig:ContractingDumbbell} in the neighbourhood of the point at which pinch-off occurs, $z=z_0$.  If we set the time at which pinch-off occurs to be $t=t_0$, then we are particularly interested in the solution behaviour in the limit $t\rightarrow t_0^-$.  It proves useful to use cylindrical polar coordinates so that $\phi = \phi(\rho,z,t)$ and the interface $\partial\Omega(t)$ is denoted by $\rho = h(z,t)$.  In these coordinates, \eqref{eq:Porous1}-\eqref{eq:Porous3} become
\begin{subequations}
\begin{alignat}{3}
&\frac{\partial^2 \phi}{\partial\rho^2}+\frac{1}{\rho}\frac{\partial \phi}{\partial\rho}
+\frac{\partial^2 \phi}{\partial z^2} = 0, \qquad \qquad & \textrm{in } & \rho>h(z,t),     \label{eq:Porous1cylindrical} \\
&\frac{\p h}{\p t} +\frac{\p \phi}{\p z}\frac{\p h}{\p z} = \frac{\p \phi}{\p\rho} &\textrm{on }& \rho=h(z,t),                                    \label{eq:Porous2cylindrical} \\
&\phi = \sigma \left(\frac{1}{h\sqrt{1 + h_z^2}} - \frac{h_{zz}}{(1+h_z^2)^{3/2}} \right) & \hspace{0.2cm} \textrm{on }& \rho=h(z,t).                                        \label{eq:Porous3cylindrical}
\end{alignat}	
\end{subequations}
We seek similarity solutions to (\ref{eq:Porous1cylindrical})-(\ref{eq:Porous3cylindrical}) of the form
\begin{equation}
\phi(\rho,z,t)=(t_0-t)^\gamma \Phi(\xi,\zeta),
\quad
h(z,t)=(t_0-t)^\alpha f(\zeta),
\label{eq:similaritysolution}
\end{equation}
\begin{equation}
\xi=\frac{\rho}{(t_0-t)^\alpha},\quad \zeta=\frac{z-z_0}{(t_0-t)^\beta}.
\label{eq:similarityvariable}
\end{equation}
In these similarity variables, the moving boundary $\rho=h(z,t)$ is described by $\xi=f(\zeta)$.  By balancing the time-dependence on each side of (\ref{eq:Porous1cylindrical})-(\ref{eq:Porous3cylindrical}), these equations become
\begin{subequations}
\begin{alignat}{3}
&\frac{\partial^2 \Phi}{\partial\xi^2}+\frac{1}{\xi}\frac{\partial \Phi}{\partial\xi}
+\frac{\partial^2 \Phi}{\partial \zeta^2} = 0, \qquad \qquad & \textrm{in } & \xi>f(\zeta),     \label{eq:Porous1similarity} \\
&-\frac{1}{3}f+\frac{1}{3}\zeta\frac{\mathrm{d}f}{\mathrm{d}\zeta}
+\frac{\partial \Phi}{\partial \zeta}\frac{\mathrm{d}f}{\mathrm{d}\zeta} = \frac{\partial \Phi}{\partial\xi} &\textrm{on }& \xi=f(\zeta),\label{eq:Porous2similarity} \\
&\Phi   = \sigma \left( \frac{1}{f\sqrt{1 + f'^2}} - \frac{f''}{(1+f'^2)^{3/2}} \right) & \hspace{0.2cm} \textrm{on }& \xi=f(\zeta),\label{eq:Porous3similarity}
\end{alignat}
\label{eq:Poroussimilarity}
\end{subequations}
provided
\begin{equation}
\alpha=1/3, \quad \beta=1/3, \quad \gamma=-1/3.
\label{eq:similarityexponents}
\end{equation}
To formulate far-field conditions for $f(\zeta)$, we require that $\partial h/\partial t$ is finite away from $z=z_0$, which implies that $-f/3+\zeta f'/3\rightarrow 0$ as $\zeta\rightarrow\pm\infty$ or, alternatively,
\begin{equation}
f'\rightarrow f/\zeta \quad \zeta\rightarrow\pm \infty.
\label{eq:Porous4similarity}
\end{equation}
By rescaling $\Phi$ by a further $\sigma^{2/3}$ and $f$, $\xi$, and $\zeta$ by $\sigma^{1/3}$, the surface tension parameter $\sigma$ can be completely eliminated from \eqref{eq:Porous1similarity}-\eqref{eq:Porous3similarity}, suggesting that the pinch-off behaviour is universal and independent of all parameters. In terms of our dimensional quantities, the resulting similarity transformations would become
\begin{equation}
\begin{split}
p - p_B &= -\left( \frac{\mu \varphi \gamma^2}{K} \right)^{1/3} (t_0 - t)^{-1/3} \Phi, \\ (\rho, z - z_0, h) &= \left( \frac{\gamma K}{\mu \varphi} \right)^{1/3}(\xi, \zeta, f).
\end{split}
\end{equation}
We do not attempt to solve (\ref{eq:Poroussimilarity}) and (\ref{eq:Porous4similarity}), but instead proceed to derive a more tractable system for the similarity solution below.  For now it is worth emphasising that we are able to derive the similarity exponents (\ref{eq:similarityexponents}) without any further analysis, which implies that (\ref{eq:similaritysolution}) is a similarity solution of the first kind.  Further, it is interesting to note the scaling exponent $\alpha = 1/3$ was observed experimentally in \cite{Ishiguro2007} when $^{3}$He crystals pinched off; we discuss this issue further in Sec.~\ref{sec:discussion}.

\subsubsection{Similarity solution via an integral equation} \label{sec:SlenderBody}

In the following our approach is similar to that performed in Eggers and coworkers~\cite{Eggers2007,Fontelos2011} who study the break up of a bubble immersed in a fluid of low viscosity.  In cylindrical polar coordinates, a solution to Laplace's equation \eqref{eq:Porous1cylindrical} can be represented as the integral
\begin{align} \label{eq:IntegralRepresentation}
	\phi(\rho,z,t) = \int_{z_0-L}^{z_0+L} \frac{C(\hat{z},t)}{\sqrt{(\hat{z}-z)^2 + \rho^2}} \, \text{d}\hat{z},
\end{align}
where the bubble has a length of $2L(t)$ and $C(z,t)$ is an unknown function to be determined. The kinematic boundary condition \eqref{eq:Porous2cylindrical} becomes
\begin{equation}
\begin{split}
\frac{\partial h}{\partial t}-\frac{\partial h}{\partial z}
\int_{z_0-L}^{z_0+L} \frac{(z-\hat{z})C(z',t)}{((\hat{z}-z)^2 + h(z,t)^2)^{3/2}} \, \text{d}\hat{z}
= \\
-h\int_{z_0-L}^{z_0+L} \frac{C(\hat{z},t)}{((\hat{z}-z)^2 + h(z,t)^2)^{3/2}} \, \text{d}\hat{z},
\end{split}
\label{eq:integralkinetic}
\end{equation}
while the dynamic boundary condition \eqref{eq:Porous3cylindrical} becomes
\begin{equation}
\begin{split}
\int_{z_0-L}^{z_0+L} \frac{C(\hat{z},t)}{\sqrt{(\hat{z}-z)^2 + h(z,t)^2}} \, \text{d}\hat{z}
= \\
\frac{1}{h\sqrt{1 + h_z^2}} - \frac{h_{zz}}{(1+h_z^2)^{3/2}}.
\end{split}
\label{eq:integraldynamic}
\end{equation}
Equations (\ref{eq:integralkinetic})-(\ref{eq:integraldynamic}), which form a closed form system for $h(z,t)$ and $C(z,t)$, are an analogue of the system derived in \cite{Eggers2007,Fontelos2011} for a bubble in a inviscid fluid.

As in Sec.~\ref{sec:similaritysoln}, we write $h$ out as a similarity solution as in (\ref{eq:similaritysolution})-(\ref{eq:similarityvariable}), where again $t=t_0$ is the pinch-off time.  We also write out  $C(z,t)=(t_0-t)^\gamma D(\zeta)$.  By balancing exponents, we find that provided $\alpha$, $\beta$ and $\gamma$ are given by (\ref{eq:similarityexponents}), then (\ref{eq:integralkinetic})-(\ref{eq:integraldynamic}) become
\begin{equation}
\frac{1}{3}(-f+\zeta f')
=\int_{-\infty}^{\infty} \frac{(f'(\zeta)(\zeta-\hat{\zeta})-f(\zeta))
D(\hat{\zeta}) \, }{((\hat{\zeta}-\zeta)^2 + f(\zeta)^2)^{3/2}}\text{d}\hat{\zeta},
\label{eq:integralkinetic2}
\end{equation}
\begin{equation}
\begin{split}
\int_{-\infty}^{\infty} \frac{D(\hat{\zeta})}{\sqrt{(\hat{\zeta}-\zeta)^2 + f(\zeta)^2}} \, \text{d}\hat{\zeta}
= \\ \frac{1}{f\sqrt{1 + f'^2}} - \frac{f''}{(1+f'^2)^{3/2}}.
\end{split}
\label{eq:integraldynamic2}
\end{equation}
Note that for this similarity solution we have taken the limit $L/(t_0-t)^{1/3}\rightarrow\infty$.  We solve (\ref{eq:integralkinetic2})-(\ref{eq:integraldynamic2}) numerically for $f$ and $D$, as described below.

Before comparing the similarity solution from (\ref{eq:integralkinetic2})-(\ref{eq:integraldynamic2}) with our numerical solutions to the full moving boundary problem, we proceed to make an approximation.  We assume the main contribution to the integral in (\ref{eq:integralkinetic2}) is local around $\hat{\zeta}=\zeta$ and expand the integrand about this point.  To leading order we find
\begin{equation}\label{eq:DApproximation}
\frac{1}{3}(-f+\zeta f')=-\frac{D}{f}\int_{-\infty}^{\infty}\frac{\mathrm{d}\hat{\eta}}{(1+\hat{\eta}^2)^{3/2}},
\end{equation}
which suggests $D\approx f(f - \zeta f')/6$.  As a result, we have from (\ref{eq:integraldynamic2}) a single equation for $f(\zeta)$:
\begin{equation} \label{eq:IntegralEquation}
\begin{split}
\frac{1}{6} \int_{-\infty}^{\infty} \frac{f(\hat{\zeta})(f(\hat{\zeta}) - \hat{\zeta} f'(\hat{\zeta}))}{\sqrt{(\hat{\zeta}-\zeta)^2 + f(\zeta)^2}} \, \text{d} \hat{\zeta}
=  \\\frac{1}{f\sqrt{1 + f'^2}} - \frac{f''}{(1+f'^2)^{3/2}}.
\end{split}
\end{equation}
%This argument provides a kind of slender body theory \cite{Eggers2007,Fontelos2011} as we effectively ignore the term $h_zu_z$ in the kinematic condition (\ref{eq:Porous2cylindrical}) when  approximating (\ref{eq:integralkinetic2}).
In contrast to the problem of bubble pinch-off in an inviscid fluid, the lengths $\rho=h$ and $z$ in our Darcy flow problem scale the same way (both with the exponent $1/3$), so this type of analysis is not exact in the limit $t\rightarrow t_0^-$.  However, as we now show, solutions to (\ref{eq:IntegralEquation}) with (\ref{eq:Porous4similarity}) still provide a very good approximation to the pinch-off profile, which is interesting.

\begin{figure}
	\centering
	\includegraphics[width=0.80\linewidth]{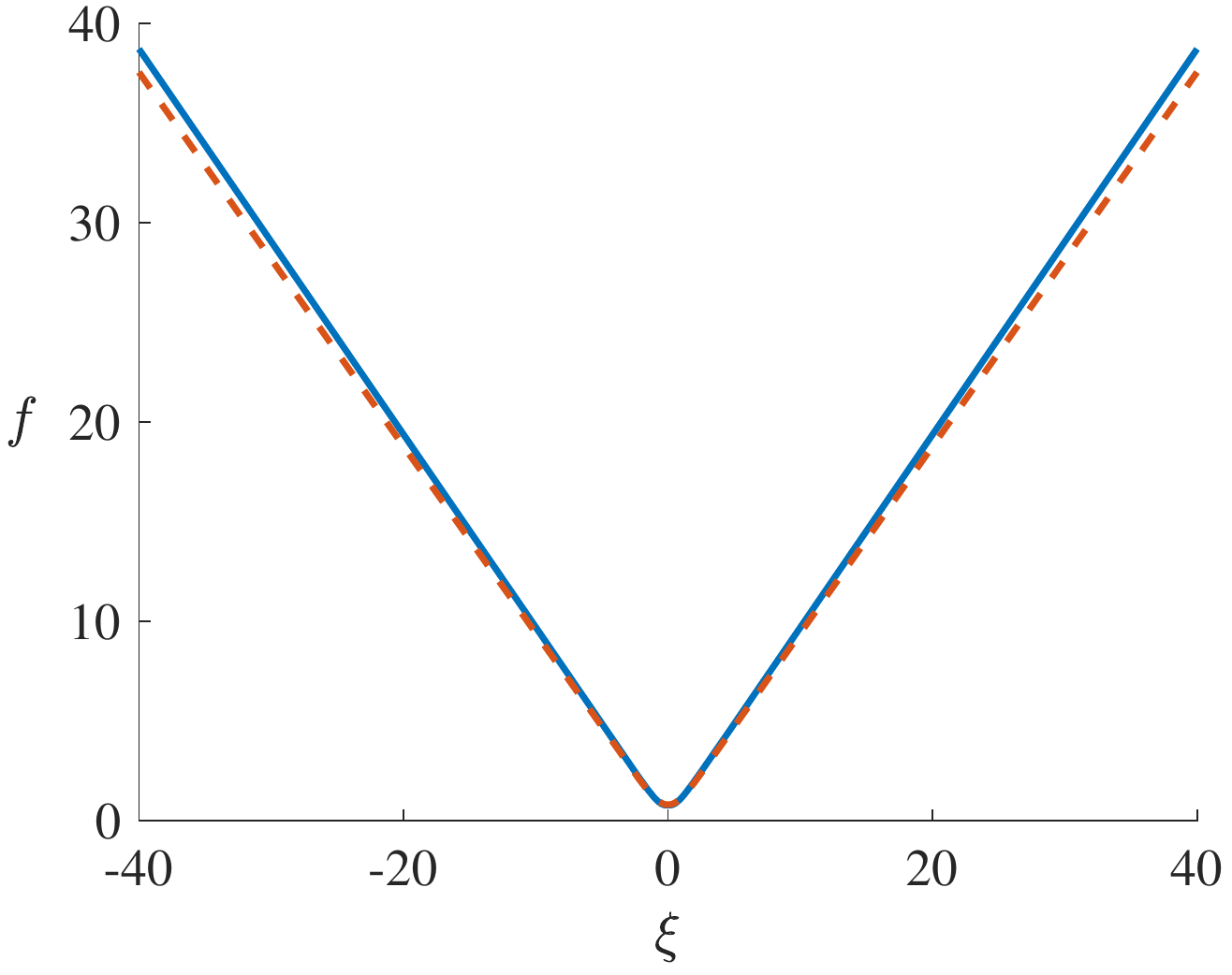}
	\caption{A comparison of numerical solution to \eqref{eq:integralkinetic2} and \eqref{eq:integraldynamic2} (solid blue) with the numerical solution to \eqref{eq:IntegralEquation} (dashed red).}
	\label{fig:selfsimilarcomparison}
\end{figure}

We compare the numerical solution of \eqref{eq:integralkinetic2}-\eqref{eq:integraldynamic2} with the numerical solution to \eqref{eq:IntegralEquation}, shown in Fig.~\ref{fig:selfsimilarcomparison}.  Solutions are computed on the truncated domain $-40 \le \xi \le 40$, which is discretised into equally spaced grid points with $\Delta \xi = 0.1$. The derivatives and integrals are approximated using central differences and the trapezoidal rule, respectively. This approach leads to a nonlinear system of algebraic equations which is solved using the Newton-Raphson method. We find that the gradient of $f$ from the solution to \eqref{eq:integralkinetic2}-\eqref{eq:integraldynamic2} is slightly steeper than the solution to \eqref{eq:IntegralEquation} in the far field. However, around the pinch-region where $\xi = 0$, the two solutions are indistinguishable (at this scale). This comparison suggests that the solution to \eqref{eq:IntegralEquation} works effectively as well as the solution to \eqref{eq:integralkinetic2} and \eqref{eq:integraldynamic2} for describing the behaviour of bubbles approaching a pinch-off event.  This observation is interesting since the spatial rescaling in the $\rho$ and $z$-directions is the same, suggesting that the neck is not slender except very close to point at which $\partial h/\partial z=0$.

The farfield boundary condition \eqref{eq:Porous4similarity} implies the linear behaviour $f \sim \pm A \xi$ as $\xi \to \pm \infty$ for some unknown $A$ such that the free surface takes the shape of a double cone at the point of pinch-off. We approximate $A$ from the numerical solution to \eqref{eq:integralkinetic2} and \eqref{eq:integraldynamic2} to be $A \approx 0.97$ which gives a cone angle of $\theta \approx 44^\circ$. In comparison, the numerical solution to \eqref{eq:IntegralEquation} gives $A \approx 0.94$ resulting in a slightly smaller angle of $\theta = 43^\circ$. We note that this cone angle is larger than the value determined experimentally by \citet{Ishiguro2007}, who approximated this angle from a video frame of the experiment just after pinch-off had occurred and found $\theta \approx 35^\circ$.

\subsubsection{Numerical solution to full moving boundary problem \label{sec:ApprochingPinchOff}}

\begin{figure}
	\centering
	\includegraphics[width=0.80\linewidth]{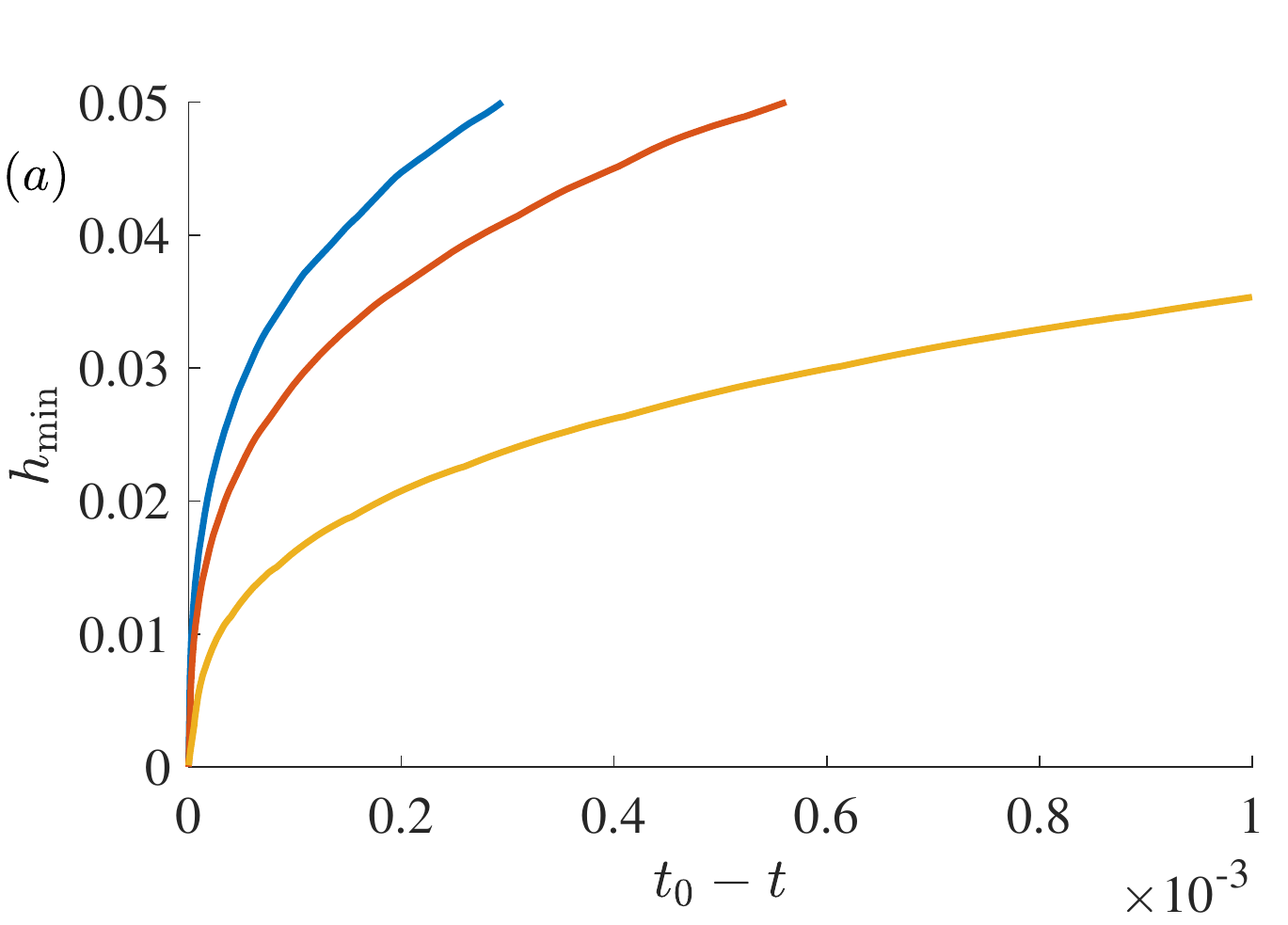} \\
	\includegraphics[width=0.80\linewidth]{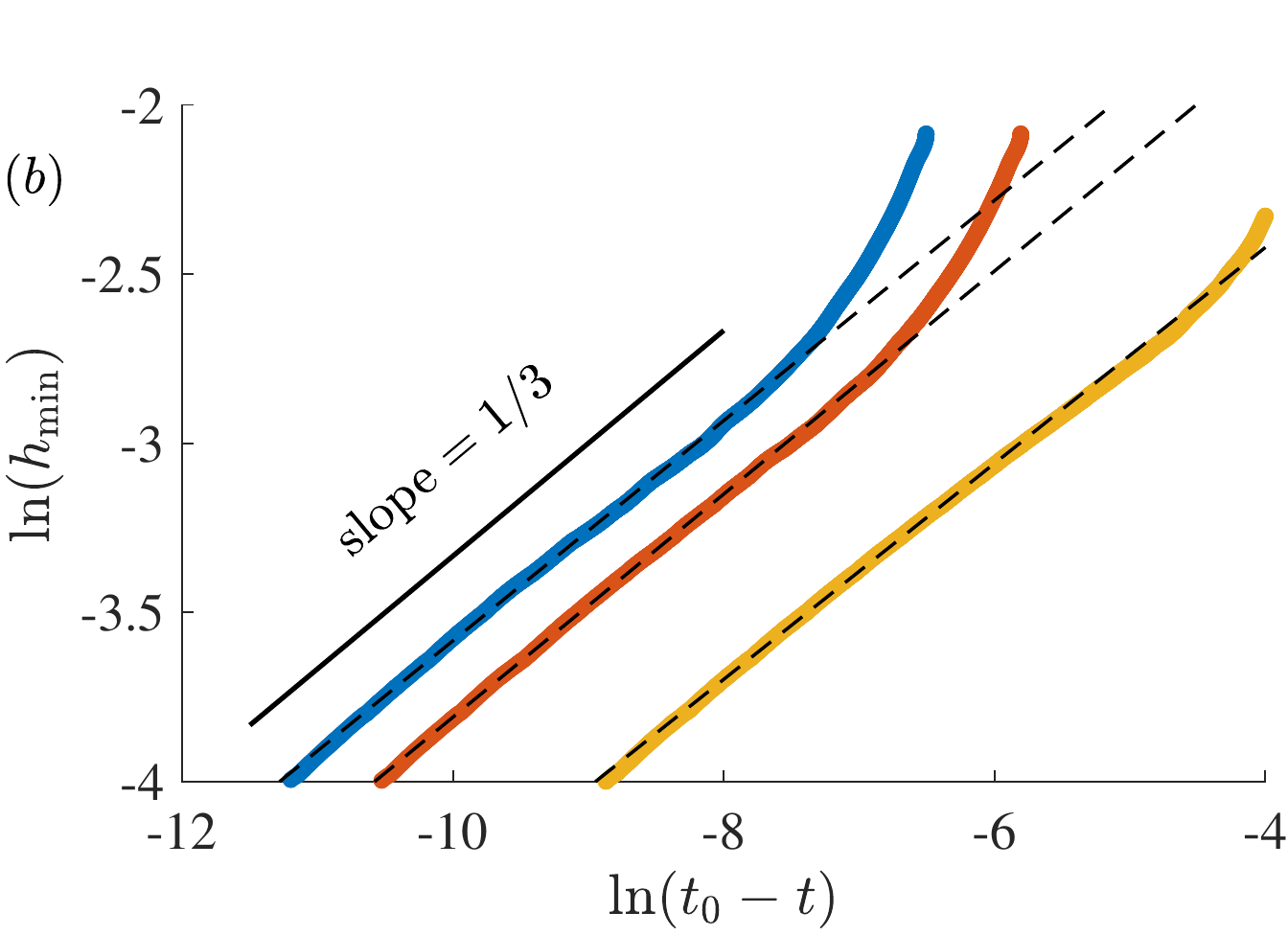}
	\caption{$(a)$ The minimum neck radius computed from the numerical solution to \eqref{eq:Spherical1}-\eqref{eq:Spherical4} as a function of time to pinch-off with, from left to right, $\sigma = 1$ (blue), 0.5 (red), and $0.1$ (yellow). $(b)$ The corresponding $\log$-$\log$ plot, where dashed curves are lines of best fit, demonstrating that our numerical scheme supports the assertion that the similarity exponent $\alpha=1/3$.}
	\label{fig:PrePinchOff}
\end{figure}

We now present numerical solutions to the full moving boundary problem \eqref{eq:Porous1}-\eqref{eq:Porous4}, produced by our numerical scheme described in Sec.~\ref{sec:NumericalScheme}.  We begin by checking whether the similarity exponent of $\alpha = 1/3$ is consistent with the full solution.  To do so, we solve \eqref{eq:Spherical1}-\eqref{eq:Spherical4} for the case in which the interface is initially an asymmetric dumbbell (see Fig.~\ref{fig:ContractingDumbbell}) for different values of the surface tension parameter $\sigma$. While increasing $\sigma$ decreases the time $t_0$ at which pinch-off occurs, the similarity exponent $\alpha$ is independent of the parameters of the model. Fig.~\ref{fig:PrePinchOff}$(a)$ compares the minimum neck radius of each of these solutions solution as a function of the time to pinch-off, $t_0 - t$, where $\sigma = 1$, (blue), 0.5 (red), and 0.1 (yellow), while Fig.~\ref{fig:PrePinchOff}$(b)$ shows the corresponding $\log$-$\log$ plot. By taking a line of best fit (black dashed lines) for each value of $\sigma$, we compute $\alpha = 0.33$ to two decimal places, which is in good agreement with the theoretical result $\alpha=1/3$ derived in Sec.~\ref{sec:similaritysoln}.

\begin{figure}
	\centering
	\includegraphics[width=0.80\linewidth]{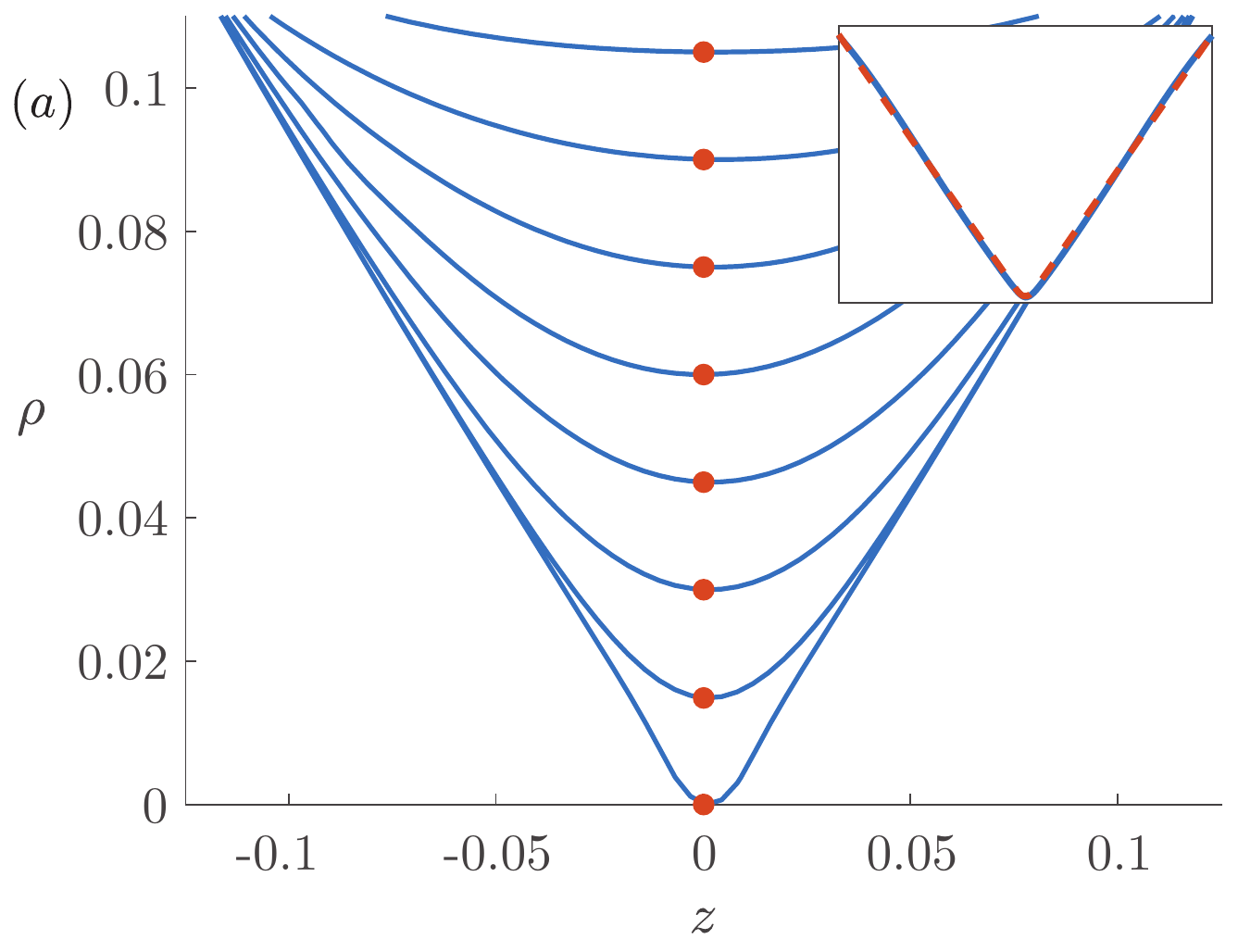} \\
	\includegraphics[width=0.80\linewidth]{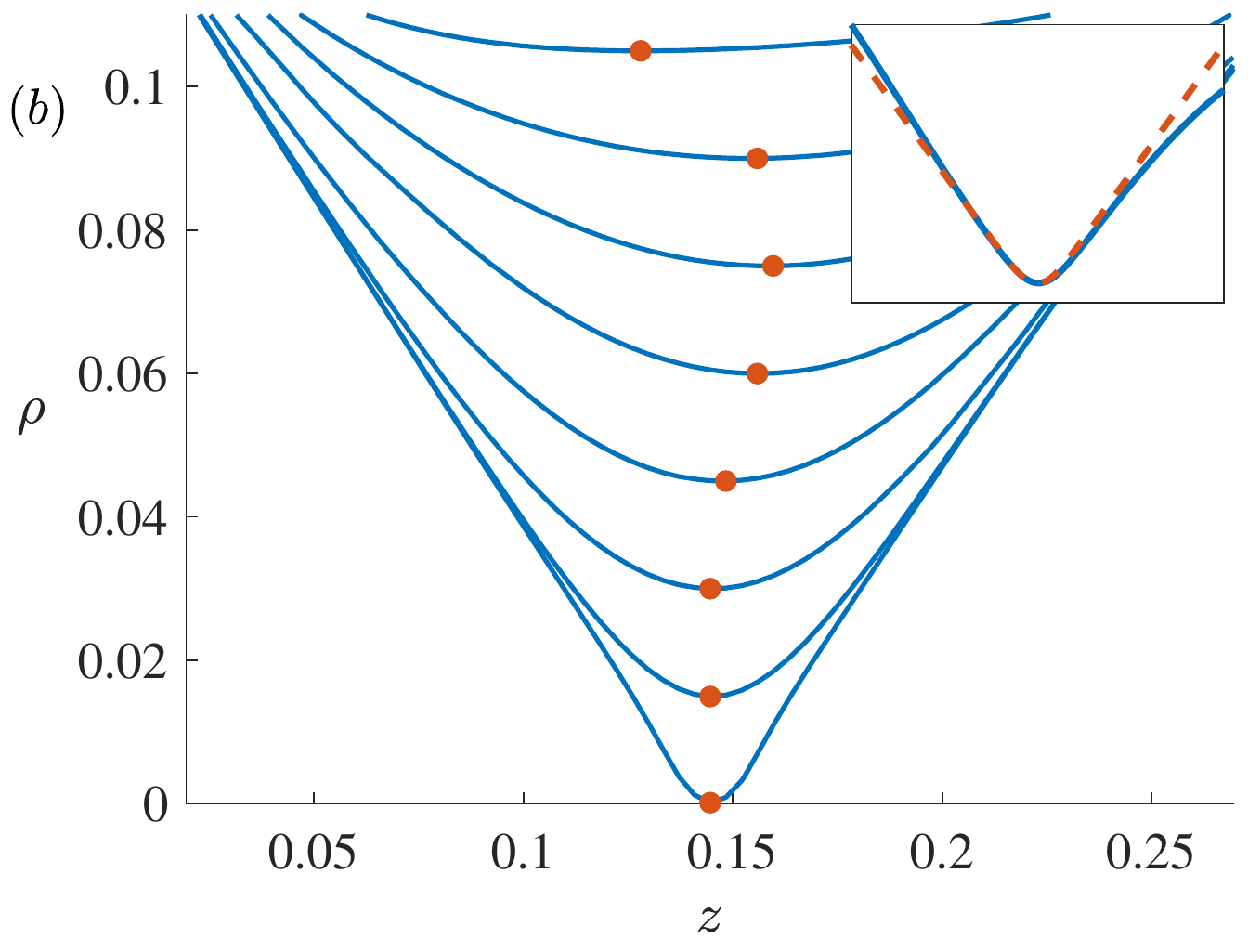}
	\caption{The time evolution of the numerical solution to \eqref{eq:Porous1}-\eqref{eq:Porous4} up to when pinch-off occurs for $(a)$ symmetric and $(b)$ asymmetric initial condition. The dots (in red) denote the location of the minimum neck radius. Inserts compare the interfacial profile rescaled according to (\ref{eq:similaritysolution})-(\ref{eq:similarityvariable}) with $\alpha = \beta = 1/3$ (solid blue) with the numerical solution to \eqref{eq:integralkinetic2} and \eqref{eq:integraldynamic2} (dashed red).}
	\label{fig:PreFullSolution}
\end{figure}

We also compare the interfacial profiles of the numerical solution to \eqref{eq:Spherical1}-\eqref{eq:Spherical4} with the similarity solution \eqref{eq:integralkinetic2}-\eqref{eq:integraldynamic2}. Fig.~\ref{fig:PreFullSolution} shows the evolution of the solution to \eqref{eq:Spherical1}-\eqref{eq:Spherical4} for both a symmetric and asymmetric initial conditions (see Fig.~\ref{fig:ContractingDumbbell}) up to $t = t_0$. The inserts compare the numerical solution of \eqref{eq:Spherical1}-\eqref{eq:Spherical4} scaled according to (\ref{eq:similaritysolution})-(\ref{eq:similarityvariable}) with the numerical solution to \eqref{eq:integralkinetic2}-\eqref{eq:integraldynamic2}. For both the symmetric and asymmetric initial conditions, around the pinch region we observe good agreement between the similarity solution and the full numerical solution. Away from the pinch region where $|\xi|$ is large, there is noticeable deviation between the numerical solution to \eqref{eq:Spherical1}-\eqref{eq:Spherical4} and the solution to \eqref{eq:integralkinetic2}-\eqref{eq:integraldynamic2}, at least for the asymmetric case. As \eqref{eq:integralkinetic2}-\eqref{eq:integraldynamic2} was derived in the limit that pinch-off is approached, these deviations are to be expected.

\subsection{Recoil} \label{sec:RadialPinchOff}

In Sec.~\ref{sec:Contracting}, we investigated the behaviour of the bubble whose boundary's velocity blows up as it approaches pinch-off due to the unbounded increase in interfacial curvature. Once pinch-off has occurred, the two newly formed satellite bubbles will rapidly recoil due to the large local curvature at each of the tips. While many studies concerned with the pinch-off of viscous or inviscid fluids have focused on the behaviour of the interface approaching pinch-off, fewer studies consider the behaviour of post break-up \cite{Eggers2014,Papageorgiou1995,Sierou2004}.  As a starting point, we expect the argument in Sec.~\ref{sec:similaritysoln} to still hold post break-up, leading to the same similarity exponents, namely $\alpha=1/3$ and $\beta=1/3$, as was computed from the pre pinch-off solution.

\begin{figure}
	\centering
	\includegraphics[width=0.80\linewidth]{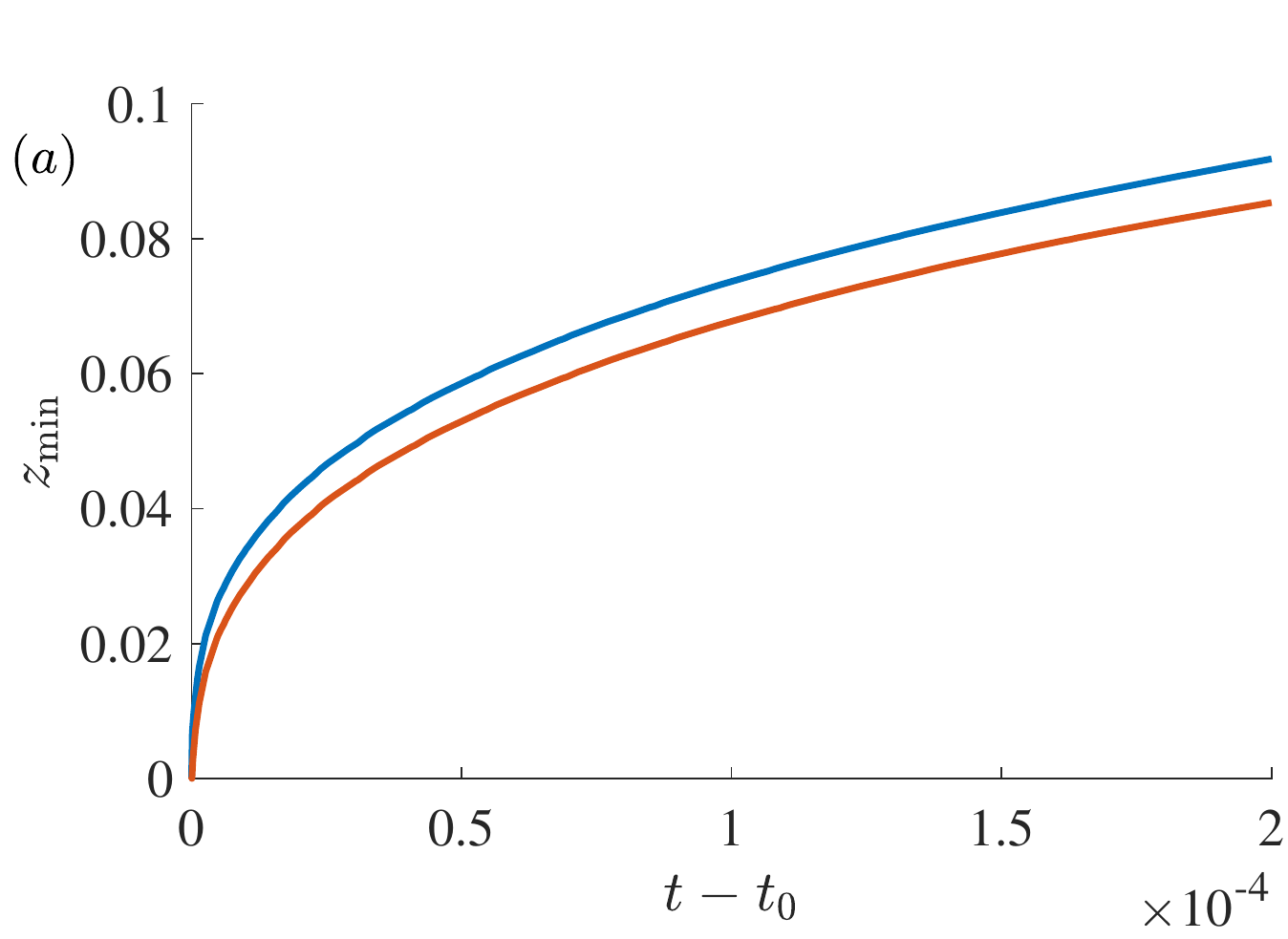}\\
	\includegraphics[width=0.80\linewidth]{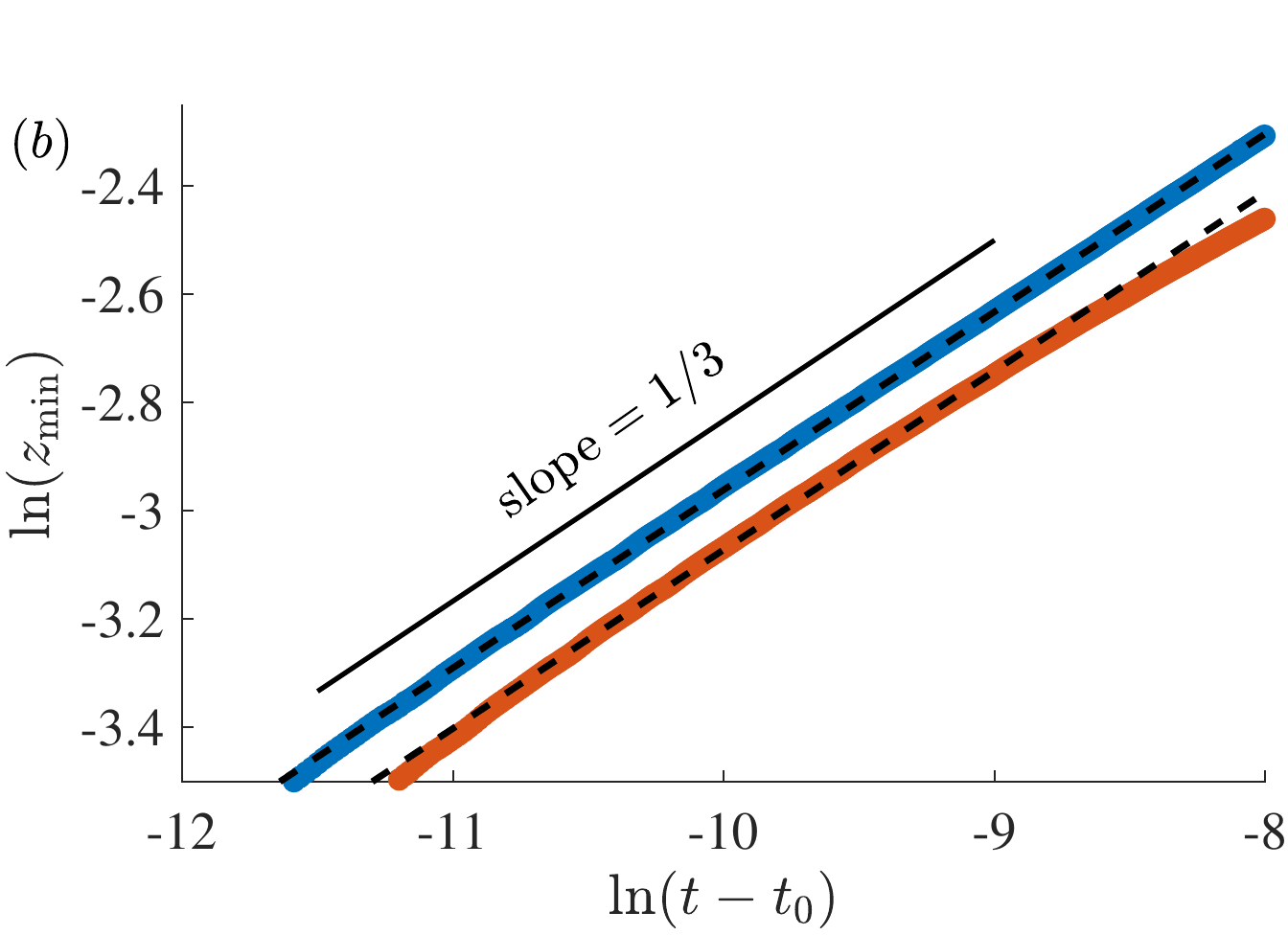}
	\caption{$(a)$ The distance between the point at which pinch-off occurs and the tip of the right (blue) and left (red) satellite bubbles. $(b)$ The corresponding $\log$-$\log$ plot, where the dashed black lines are a line of best fit, providing evidence that our numerical solutions post-pinch-off scale like (\ref{eq:SelfSimilarPost}).}
	\label{fig:PostPinchOff}
\end{figure}

To confirm that our numerical solution produces the scaling exponent of $\beta = 1/3$, we consider the evolution of an asymmetric dumbbell bubble (see Fig.~\ref{fig:ContractingDumbbell}) for $t>t_0$. After pinch-off, we compute the distance between the point where pinch-off occurs and the tip of the two retracting satellite bubbles as a function of the time. Fig.~\ref{fig:PostPinchOff}$(a)$ shows the evolution of this distance for both the left and right satellite bubbles, while Fig.~\ref{fig:PostPinchOff}$(b)$ is the corresponding plot on a $\log$-$\log$ scale. By taking a line of best fit, we find an approximation of the similarity exponent to be $\beta = 0.33$ to two decimal plates for both bubbles, which is consistent with the theoretical prediction. We repeat this procedure for a range of values of $\sigma$ (not shown) and, as was found in Sec.~\ref{sec:ApprochingPinchOff}, our numerical results are in good agreement with the theoretical prediction of $\beta = 1/3$.

\begin{figure}
	\centering
	\includegraphics[width=0.80\linewidth]{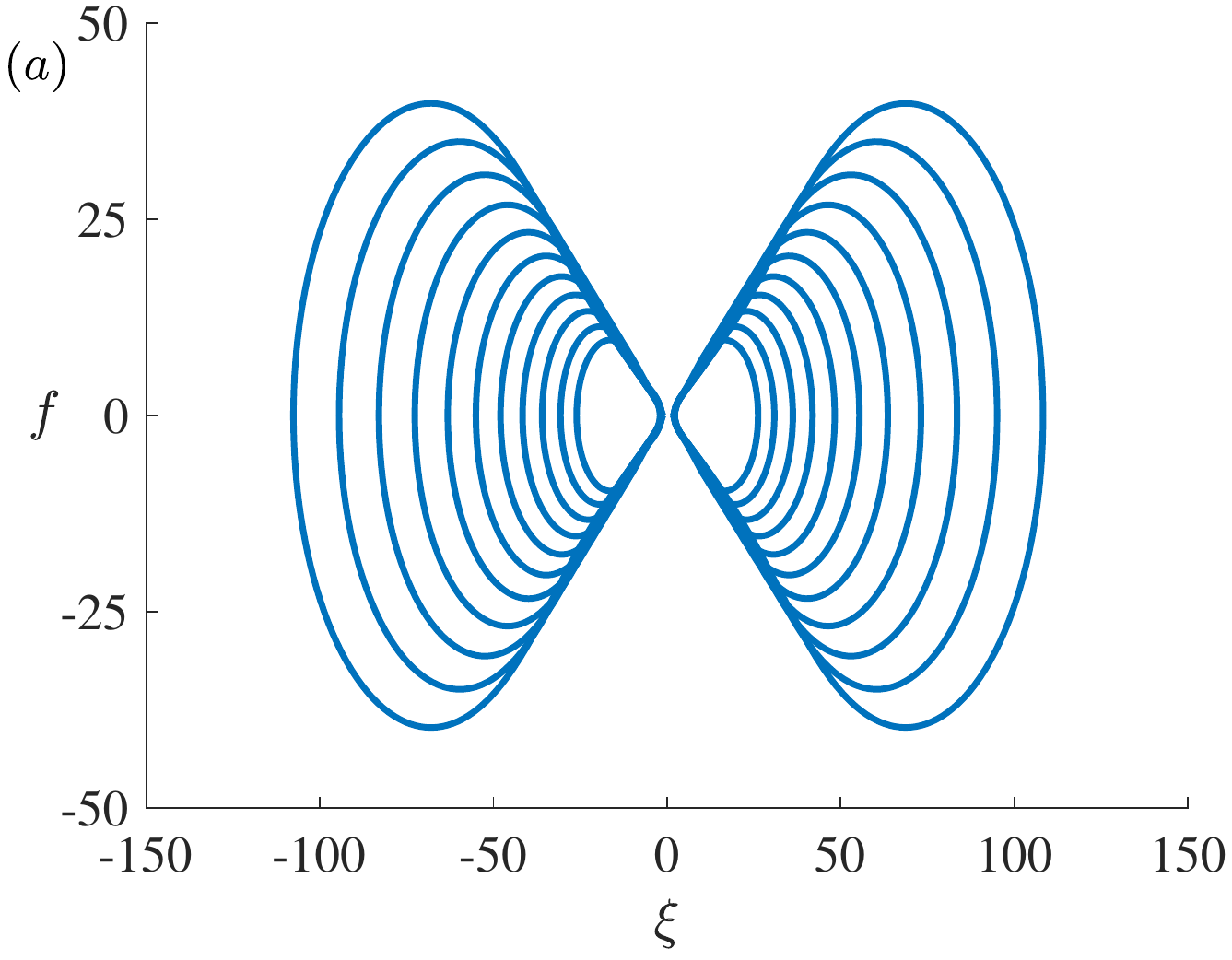}
	\includegraphics[width=0.80\linewidth]{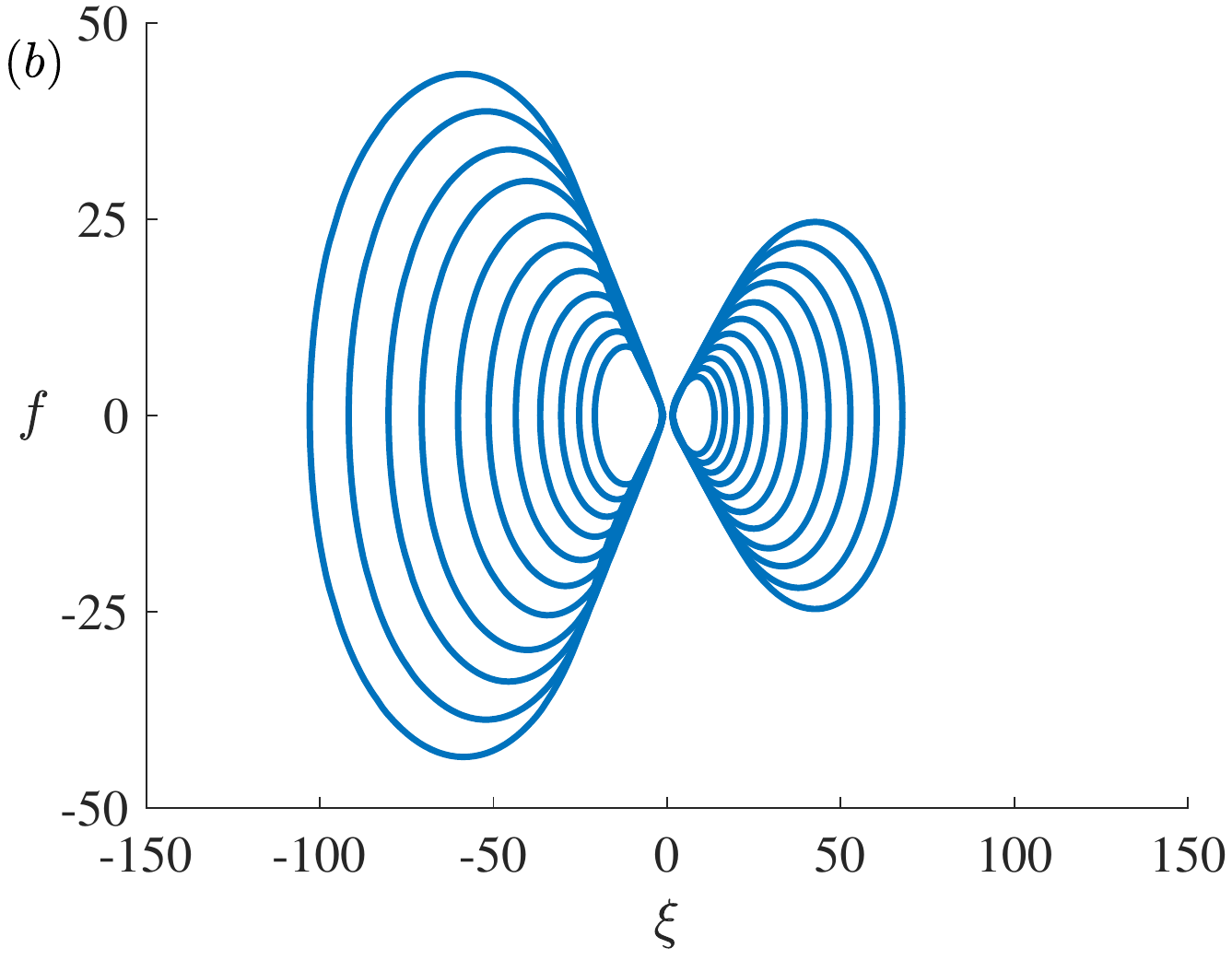}
	\caption{The numerical solution to \eqref{eq:Spherical1}-\eqref{eq:Spherical4} scaled according to \eqref{eq:SelfSimilarPost} for $(a)$ symmetric and $(b)$ asymmetric dumbbell initial condition.  These solutions illustrate how the full numerical solution post-pinch-off approaches a similarity solution for $|f|\ll 1$.}
	\label{fig:PostPinchScaled}
\end{figure}

Just after pinch off, we expect the interface to also act in a self-similar fashion.  In this case, however, the representation \eqref{eq:IntegralRepresentation} is unlikely to be valid \citep{Fontelos2011}, and the approximation \eqref{eq:DApproximation} no longer true, as $f'(\xi) \to \infty$ where the interface intersects the $\xi$-axis. Due to this difficulty we do not solve the similarity problem here. While we do not provide a numerical solution of the asymptotic equations post break-up, we perform a numerical test to confirm that solutions to \eqref{eq:Spherical1}-\eqref{eq:Spherical4} are self-similar in the limit.  We redefine our self-similar solution after pinch-off to be
\begin{align} \label{eq:SelfSimilarPost}
h = (t-t_0)^{1/3} f \left( \xi \right) \quad \text{where} \quad \xi = \frac{z-z_0}{(t-t_0)^{1/3}},
\end{align}
noting that now $t > t_0$. In Fig.~\ref{fig:PostPinchScaled}, we plot the numerical solution to \eqref{eq:Spherical1}-\eqref{eq:Spherical4} scaled according to \eqref{eq:SelfSimilarPost} for times where $t - t_0 \ll 1$ for the $(a)$ symmetric and $(b)$ asymmetric dumbbell initial conditions. For both cases, Fig.~\ref{fig:PostPinchScaled} suggests that the interfaces do indeed converge to a self-similar profiles around the pinch region.

\section{Translating bubbles in a cylindrical tube\label{sec:propagating}}

We now turn our attention to a situation in which an inviscid bubble is translating from left to right in a cylindrical tube of unit diameter, as illustrated in Fig.~\ref{fig:Geometry}$(b)$-$(c)$.  In Fig.~\ref{fig:Geometry}$(b)$, the viscous fluid fills the domain to the right of the bubble interface, while in Fig.~\ref{fig:Geometry}$(c)$ the bubble is finite and the surrounding viscous fluid extends infinitely far in both directions.  In cylindrical polar coordinates with the $z$-axis running down the centre of the cylinder, velocity potential is $\phi = \phi(\rho, z, t)$ and the interface is denoted by $\rho = h(z, t)$.  We impose a no-flux boundary condition on the cylinder wall so that $\p \phi/\p\rho = 0$ on $\rho = 1/2$. Further, our flow is driven by a flux of fluid in the far field so that $\p \phi/\p z \sim 1$ as $z \to \infty$.  It is straightforward to adapt the numerical scheme presented in Sec.~\ref{sec:NumericalScheme} to this new geometry.

\begin{figure}
	\centering
	\includegraphics[width=0.49\linewidth]{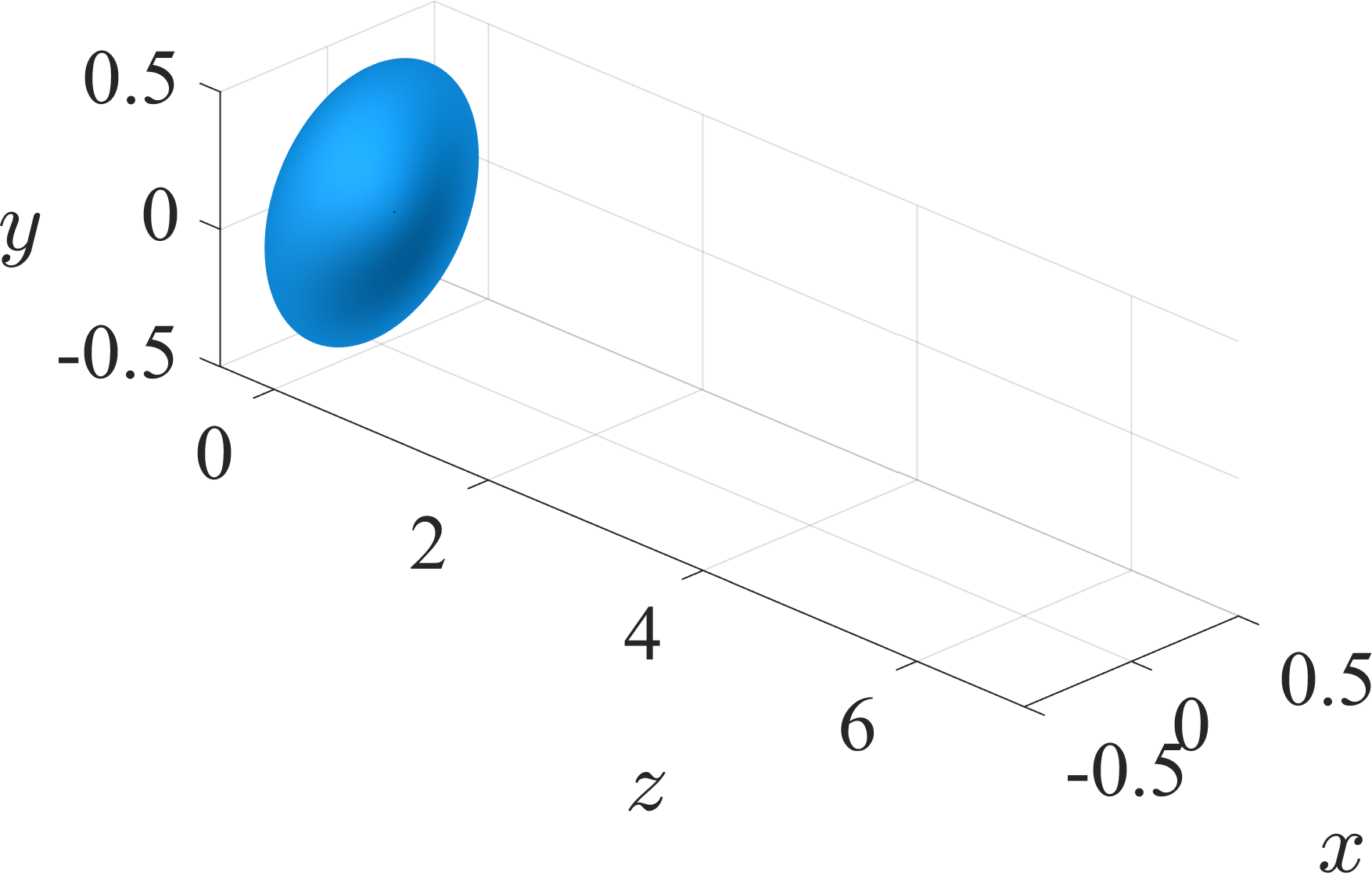}
	\includegraphics[width=0.49\linewidth]{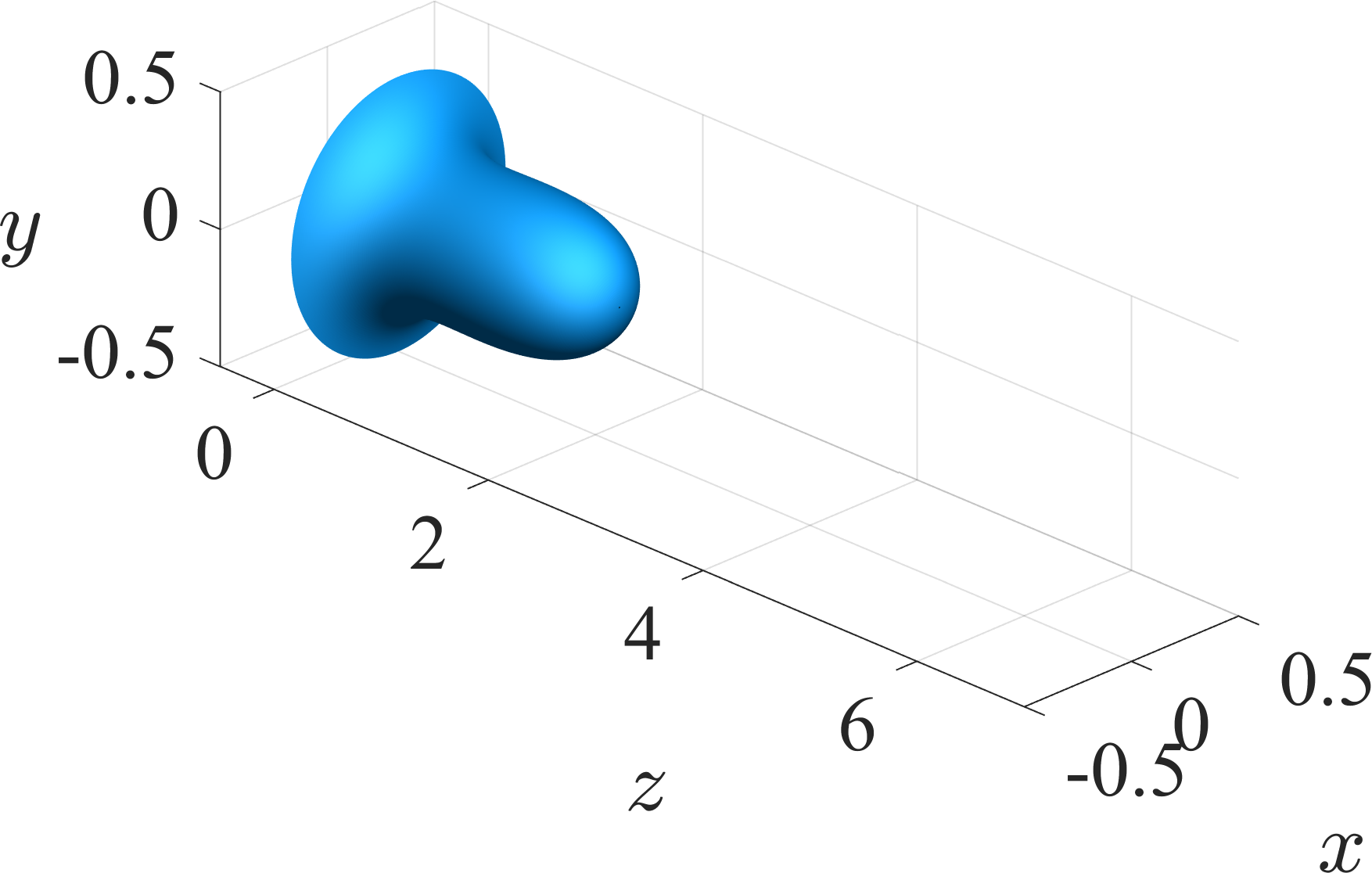}\\
	\includegraphics[width=0.49\linewidth]{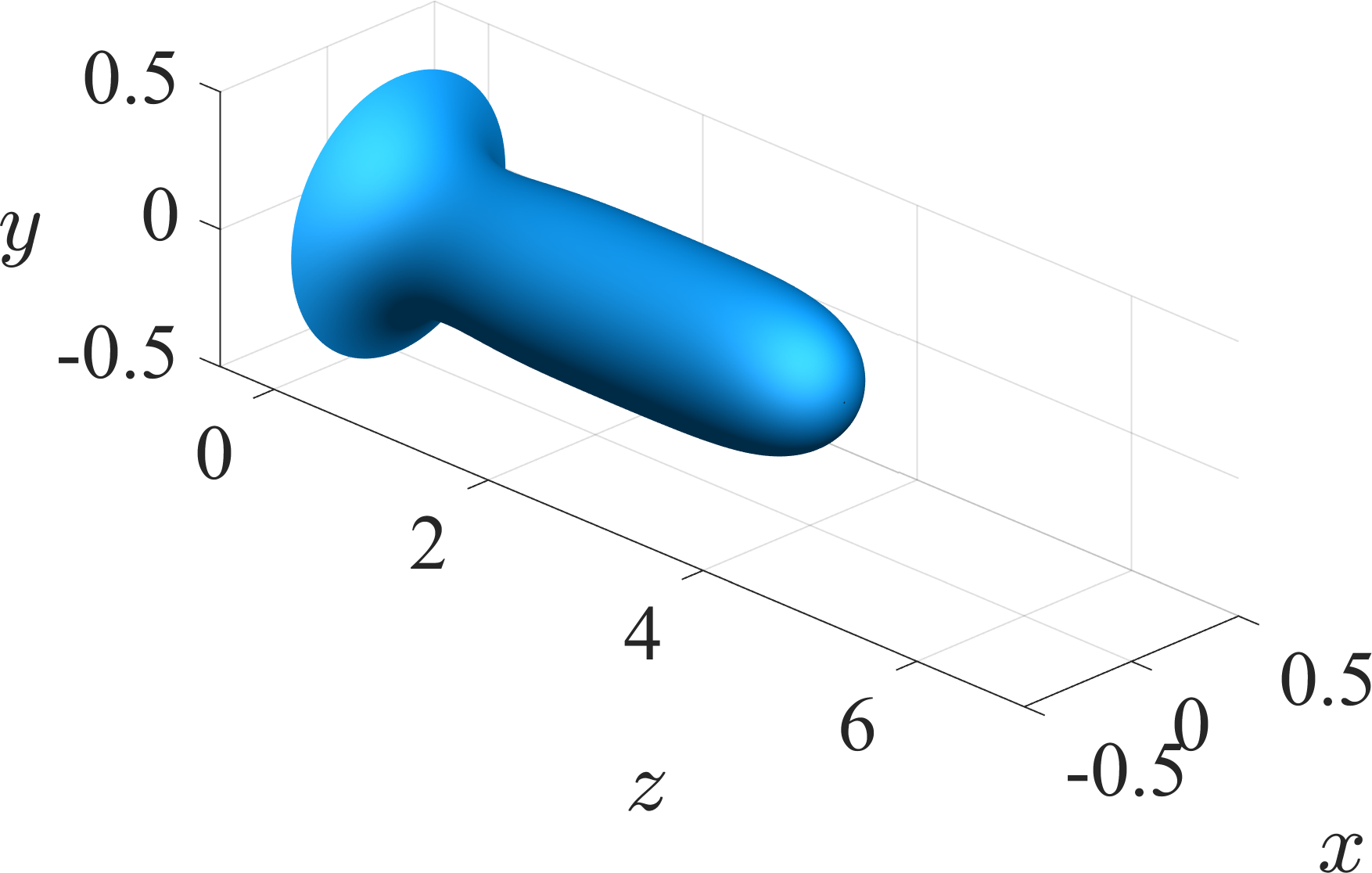}
	\includegraphics[width=0.49\linewidth]{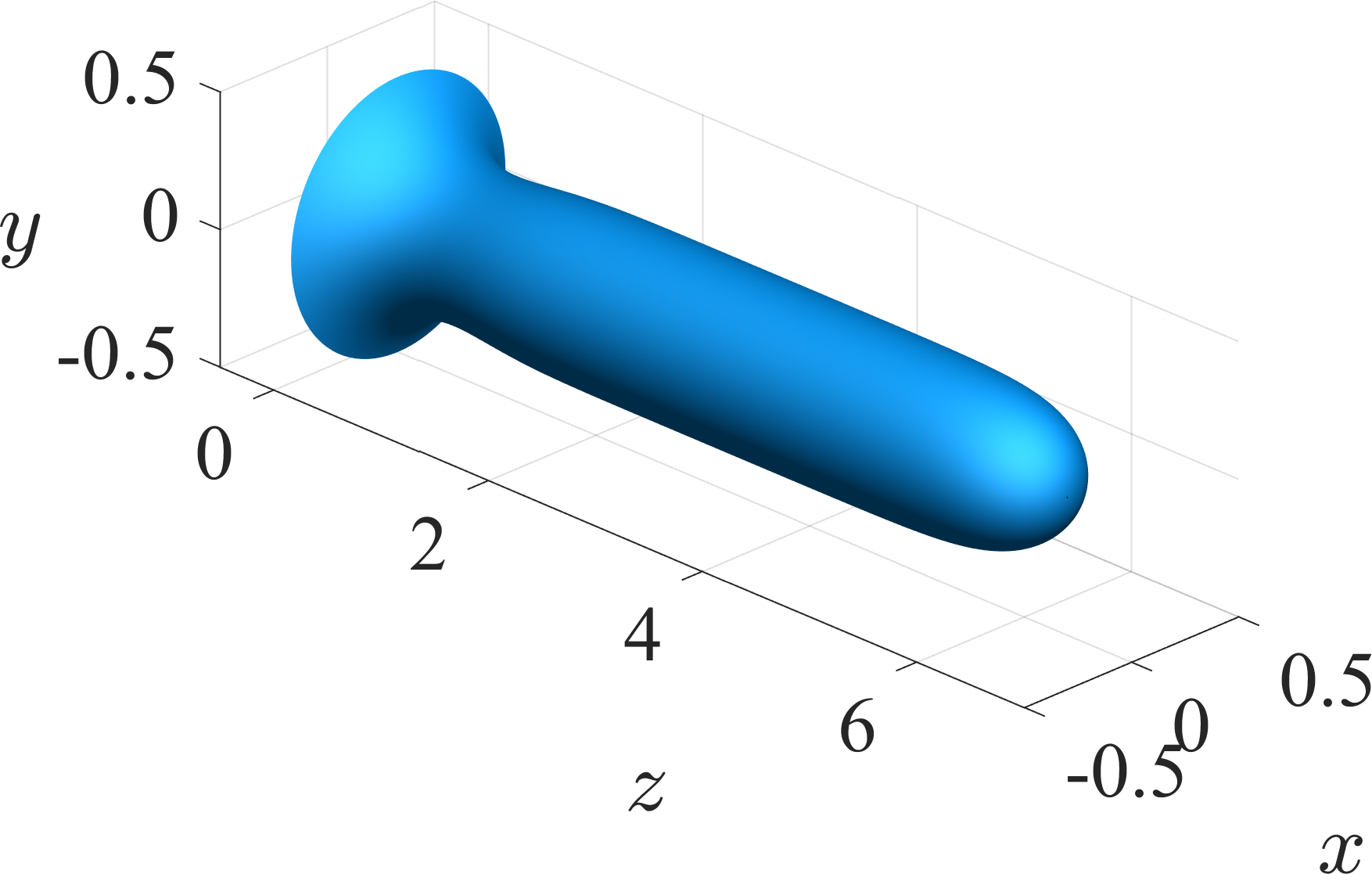}
	\caption{The evolution of a bubble in a cylindrical tube. The bubble initially has a flat interface with a small sinusoidal perturbation. As the interface grows, it becomes unstable and develops a long finger analogous to the two dimensional Saffman-Taylor finger which forms in Hele-Shaw cells. As time increases, both the speed and shape of this tip tend towards a constant.}
	\label{fig:3DSaffmanTaylor}
\end{figure}

\subsection{Development of a Saffman-Taylor-like finger}	\label{sec:SaffmanTaylorFinger}

We consider here the geometry in Fig.~\ref{fig:Geometry}$(b)$ where the inviscid fluid is displacing the viscous fluid.  The Saffman-Taylor instability causes a small perturbation of the interface to grow and eventually a single long finger of inviscid fluid will develop and propagate into the viscous fluid.  To illustrate this progression, we show in Fig.~\ref{fig:3DSaffmanTaylor} a solution for which the initial condition is a flat interface with a small sinusoidal perturbation.  The evolution to a long finger is clear.

This example is an axisymmetric analogue of the well-studied Hele-Shaw problem for which a Saffman-Taylor finger develops in a rectangular channel~\cite{Saffman1958}.  As is well known in that two-dimensional problem, both experimentally and in time-dependent numerical simulations, a single finger develops whose width $\lambda$ appears to be slightly more than half of the channel diameter provided the capillary number is sufficiently high \cite{Degregoria1986,Tabeling1987}.  Assuming a travelling wave solution with a finger moving at speed $U$, numerical studies show there is a countably infinite number of possible finger widths $\lambda$ for each fixed value of the surface tension $\sigma$ \cite{Mclean1981,Vanden1983,GardinerRIP}.  The lowest of these finger widths has been shown to correspond to a stable solution while the remaining solutions are unstable \cite{Bensimon1986}.  Each family of solutions has the asymptotic behaviours $\lambda\rightarrow 1/2^+$ as $\sigma\rightarrow 0$ and $\lambda\rightarrow 1^-$ as $\sigma\rightarrow\infty$ \cite{Chapman99,CombescotEtAl86,Hong86,Tanveer1987c}.

Returning to the axisymmetric version of this problem with which we are concerned with here, there are numerical results for the travelling wave problem by Levine \& Tu~\cite{Levine1992}, who show that much of the behaviour is qualitatively similar to the Hele-Shaw case, with multiple solutions for fixed values of the surface tension $\sigma$.  However, a key difference is that the solution branches do not all tend towards $\lambda=1/2$ as $\sigma$ decreases, but instead solutions only exist down to a minimum value of $\sigma$ at which point pairs of branches merge together (this type of merging of pairs of branches also occurs for the Hele-Shaw wedge problem studied by Ben Amar \& Combescot \cite{Amar1991,Combescot1991,Combescot1992}).

Our contribution here is to study the full time-dependent axisymmetric problem which, as far as we know, has not been documented before.  To gather data equivalent to that provided in~\cite{Levine1992}, we must compute the transient solution for a sufficient amount of time such that the finger which develops evolves at a constant shape and speed.
%Noting that the velocity of the finger is inversely proportional to the square of the finger width, this gives $\lambda = \sqrt{1 / U}$. We note that for the two dimensional case, $\lambda = 1/U$.
Of course, this approach only allows us to track the stable branch, as the solution will not evolve to an unstable travelling wave solution.

\begin{figure}
	\centering
	\includegraphics[width=0.80\linewidth]{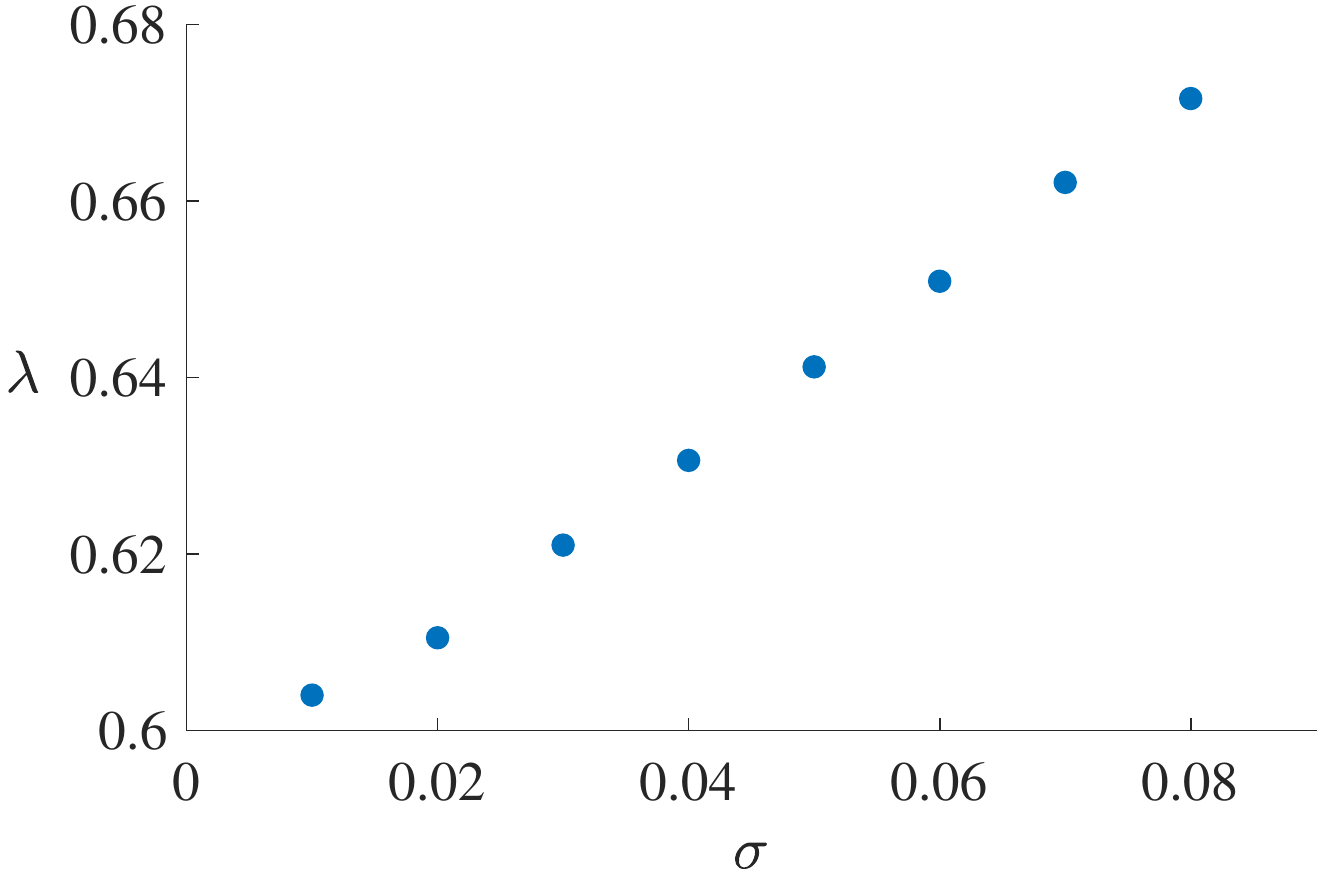}
	\caption{The finger width $\lambda$ as a function of the surface tension parameter, $\sigma$, of an axially symmetric analogue of the Saffman-Taylor finger.  The finger width is determined by evolving a near flat interface with a small sinusoidal perturbation until both the shape and speed of the finger are constant.}
	\label{fig:FingerWidth}
\end{figure}

We show in Fig.~\ref{fig:FingerWidth} the results of our calculations with various values of the finger width $\lambda$ plotted against surface tension $\sigma$.  The smallest value of finger width we observe is $\lambda \approx 0.6$, which is similar to the minimum value observed by Levine \& Tu~\cite{Levine1992}.  Our time-dependent calculations for smaller values of surface tension do not lead to single steadily propagating fingers, but instead a type of axially symmetric tip-splitting occurs, where a dimple forms at the front of the bubble and subsequently the bubble takes the shape of an off-axis finger rotated around the $z$-axis. We speculate that this bubble shape would not be observed experimentally as it is likely unstable to non-axisymmetric perturbations.

\subsection{Pinch-off of a Saffman-Taylor-like finger}	\label{sec:SaffmanTaylorFinger2}

\begin{figure}
	\centering
	\includegraphics[width=0.49\linewidth]{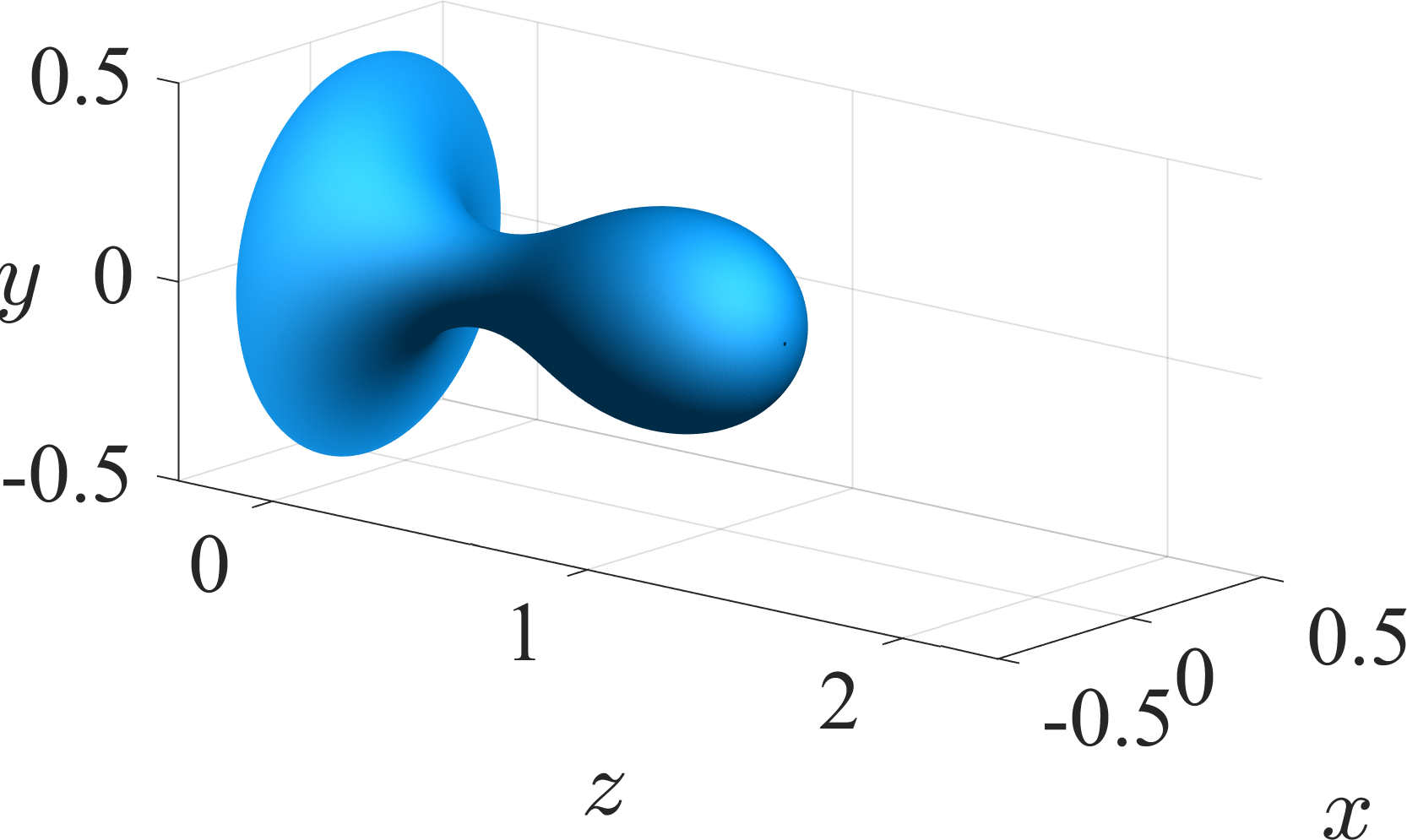}
	\includegraphics[width=0.49\linewidth]{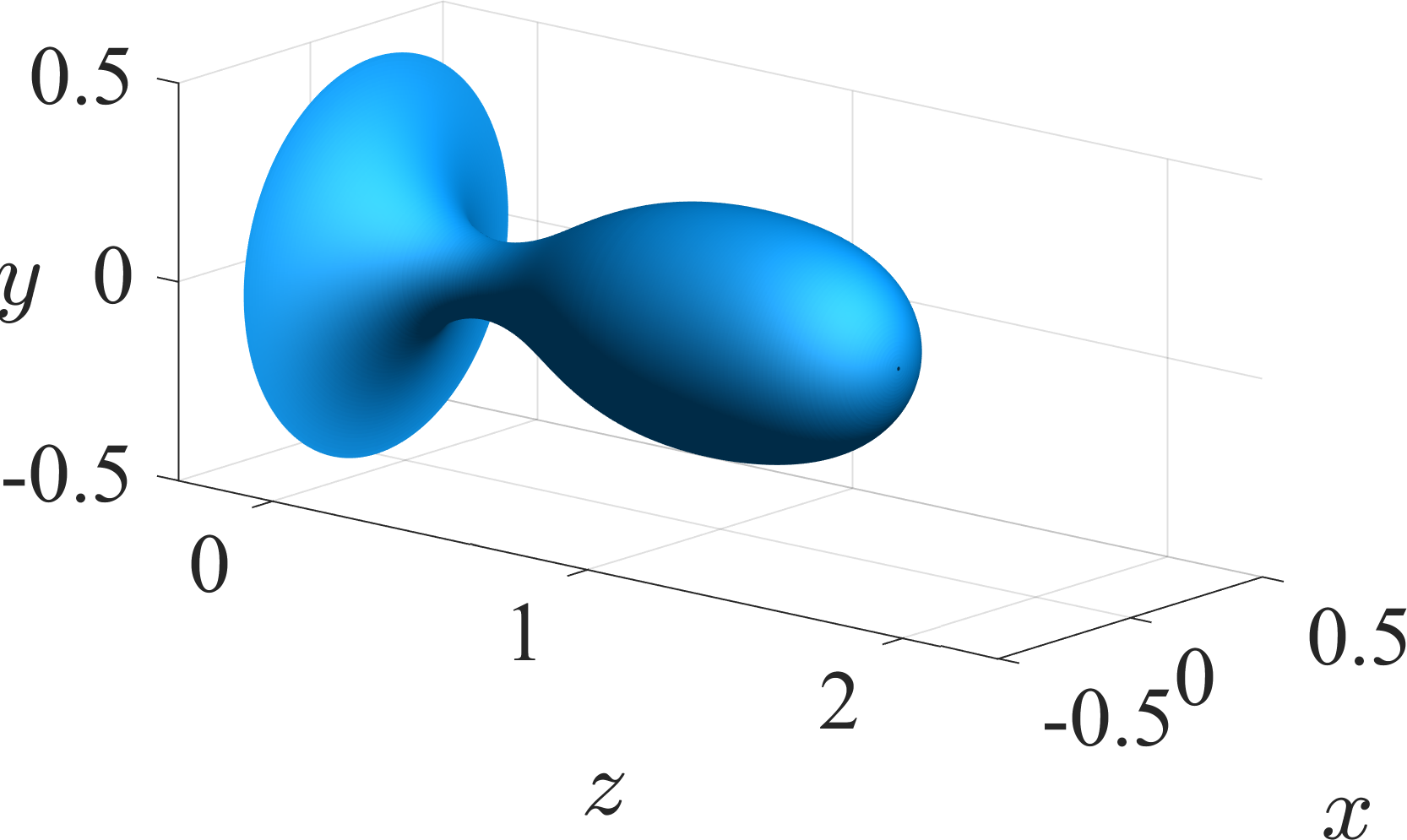}\\
	\includegraphics[width=0.49\linewidth]{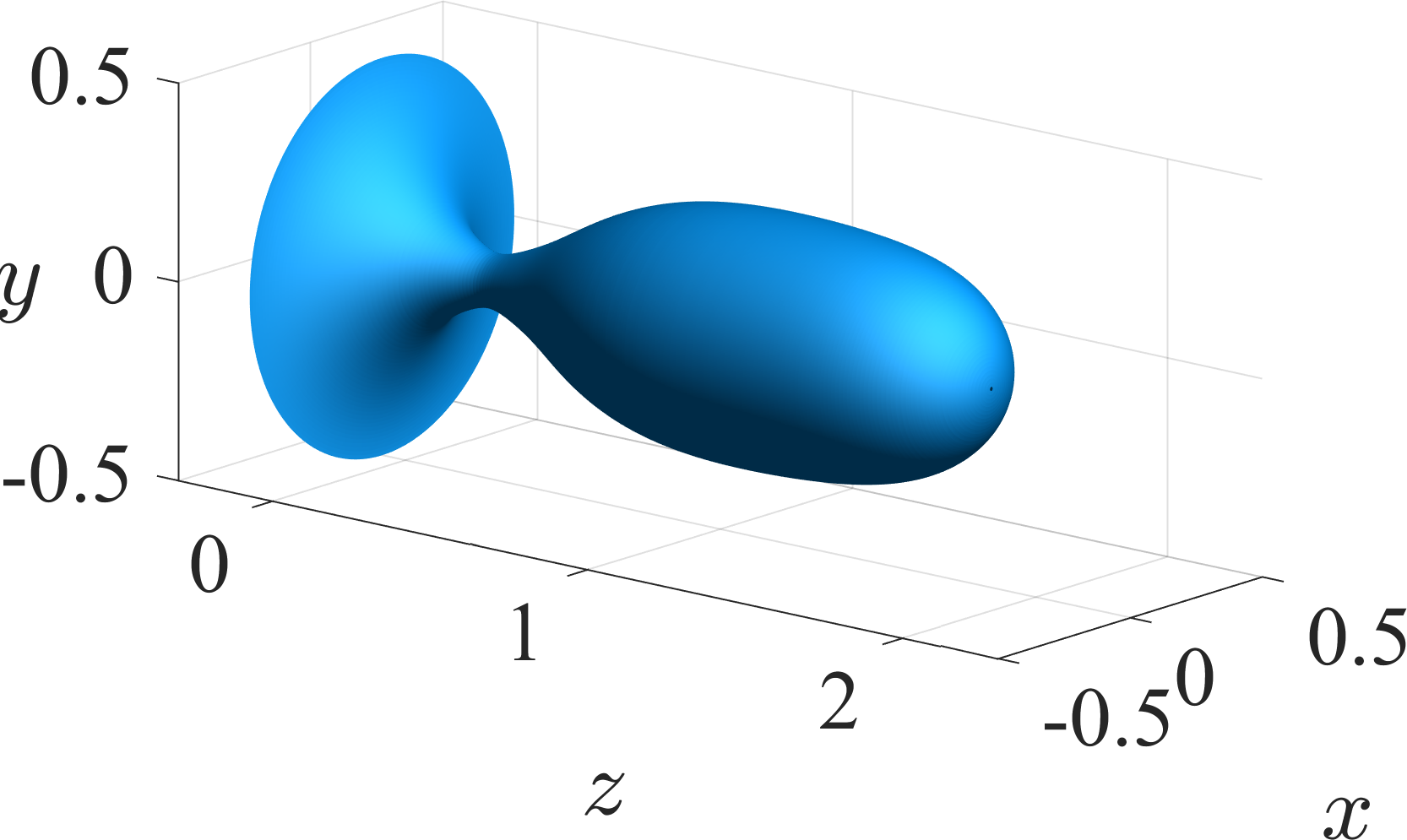}
	\includegraphics[width=0.49\linewidth]{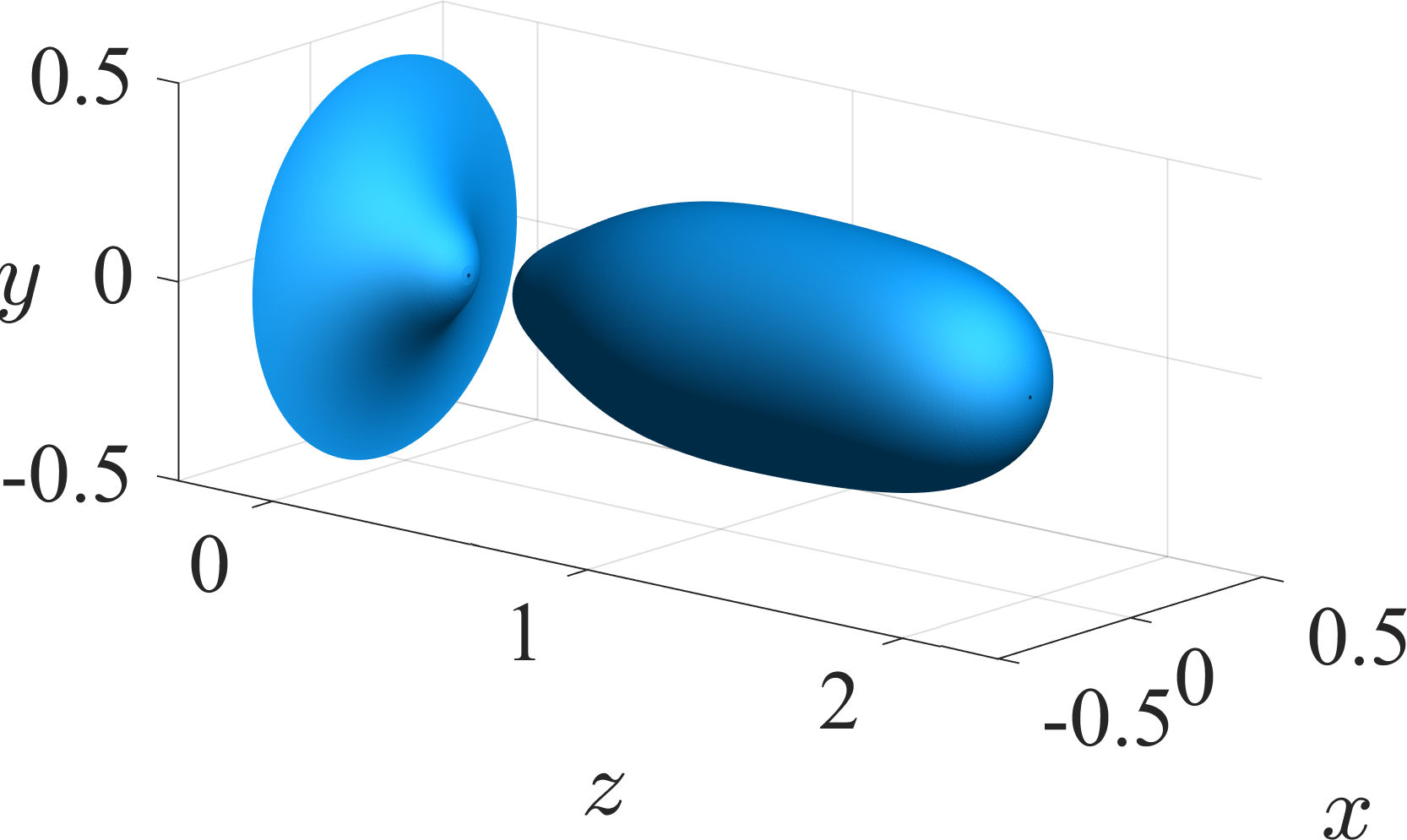}
	\caption{Numerical simulation of an expanding bubble in a cylindrical tube computed using the numerical scheme described in Sec.~\ref{sec:NumericalScheme}. As the bubble expands, the neck of the interface is small enough such that the curvature in the cylindrical direction is sufficiently large to pull the interface towards a pinch-off singularity. }
	\label{fig:expandingfinger}
\end{figure}

The simulations discussed in Sec.~\ref{sec:SaffmanTaylorFinger} were initiated with a small perturbation of flat interface.  For very different non-convex initial conditions which involve a thin neck, the bubble can undergo a pinch-off event which is similar to that described in Sec.~\ref{sec:Contracting}.  One such example is provided in Fig.~\ref{fig:expandingfinger}, illustrating how the very large azimuthal curvature in the neck region `pulls' the interface towards pinch-off even though the volume is increasing.  These singularities do not develop in the classical two-dimensional Hele-Shaw problem as there is no increasingly large component of curvature in that case.  As such, this behaviour is another example of a difference between the axially symmetric problem we consider here in this paper and the Hele-Shaw version.

\begin{figure}
	\centering
	\includegraphics[width=0.80\linewidth]{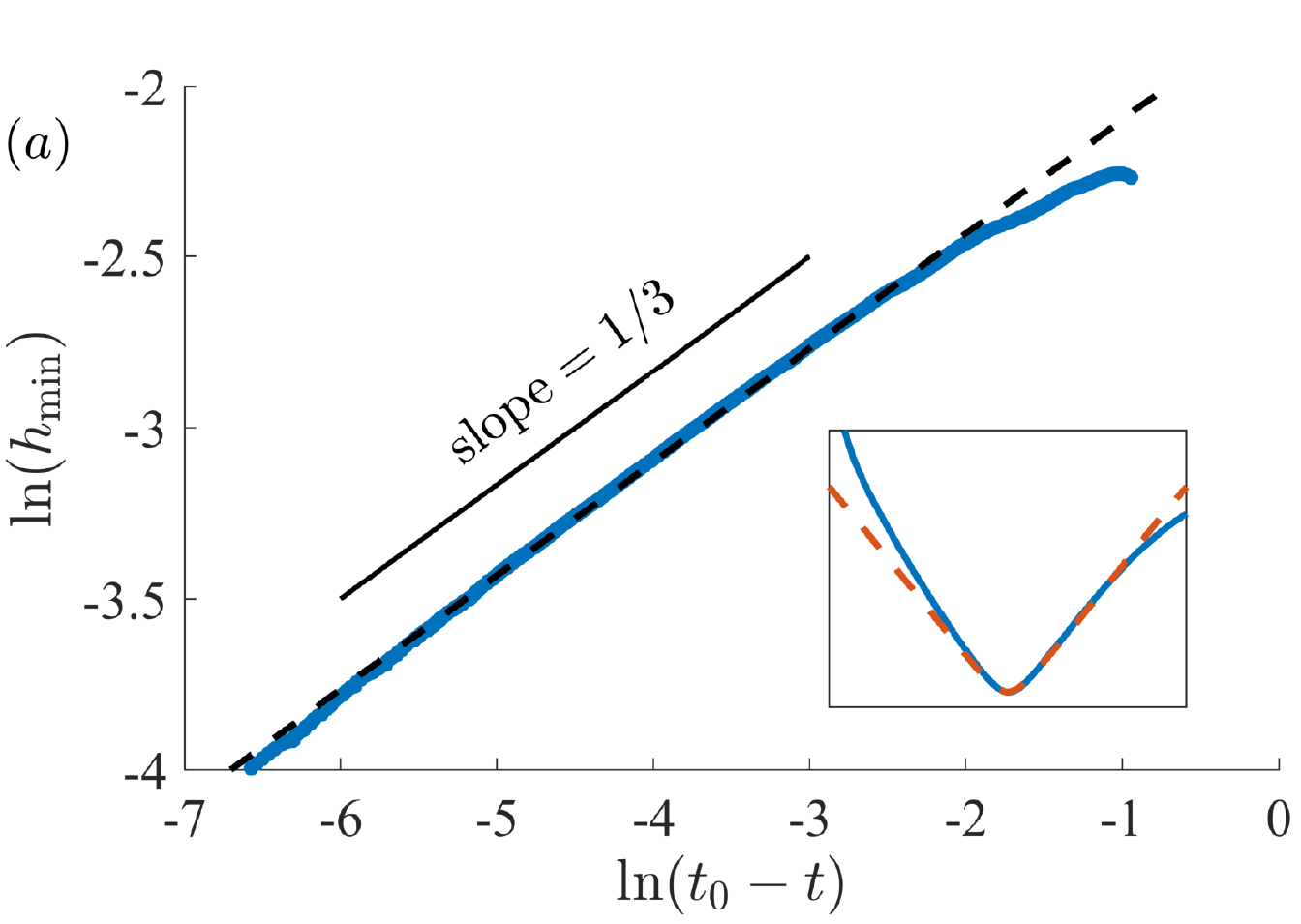} \\
	\includegraphics[width=0.80\linewidth]{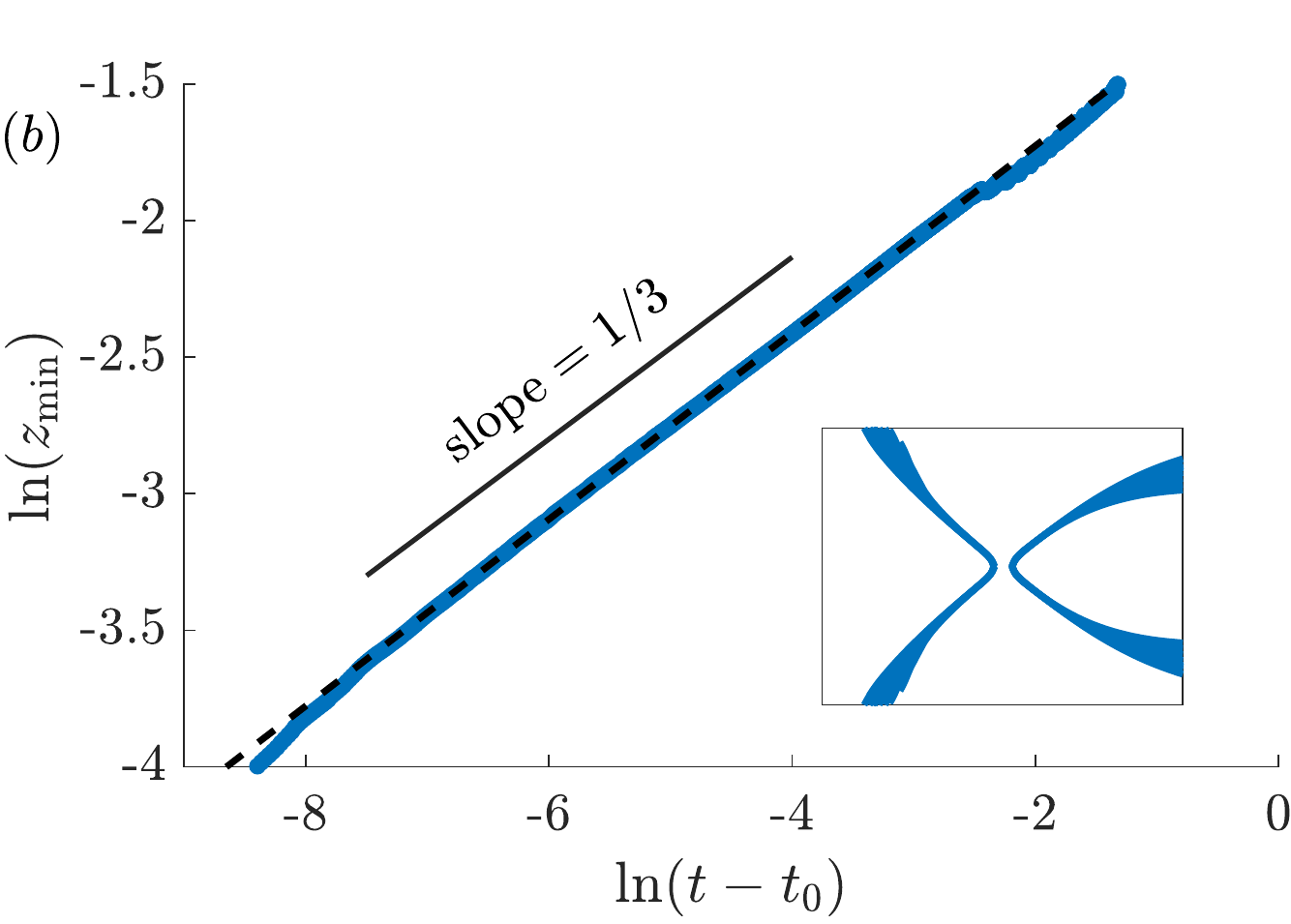}
	\caption{$(a)$ $\log$-$\log$ plot of minimum neck radius of bubble in cylindrical tube geometry, shown in Fig.~\ref{fig:expandingfinger}. The dashed curve is a line of best fit with gradient $\alpha$. Insert is comparison of solution to \eqref{eq:integralkinetic2} and \eqref{eq:integraldynamic2} (dashed red) with full numerical solution scaled according to \eqref{eq:similaritysolution} (solid blue). $(b)$ $\log$-$\log$ plot of distance between the point where pinch-off occurs and the receding tip of the bubble, compared with a line of slope $\beta=1/3$. The insert is numerical solution scaled according to \eqref{eq:similaritysolution} for times when $t - t_0 \ll 1$.}
	\label{fig:ChannelPinchOff}
\end{figure}

We now demonstrate that the pinch-off singularity that occurs in this cylindrical geometry is the same as that presented in Sec.~\ref{sec:Contracting} for a shrinking bubble.  Following the same methodology, we estimate $\alpha$ from the full numerical solution by taking a line of best fit of the minimum neck radius $h_{\mathrm{min}}$ as a function of the time to pinch-off, $t_0-t$, on a $\log$-$\log$ scale, as shown in Fig.~\ref{fig:ChannelPinchOff}$(a)$. We find a value of $\alpha = 0.33$ to two decimal places, which is again consistent with both the similarity problem derived in  Sec.~\ref{sec:similaritysoln} and the numerical solution of the contracting bubble presented in Sec.~\ref{sec:ApprochingPinchOff}. Furthermore, the insert compares the full numerical solution scaled according to \eqref{eq:similaritysolution} with the solution to \eqref{eq:integralkinetic2}-\eqref{eq:integraldynamic2}. As we observed in Fig.~\ref{fig:PreFullSolution}, around the pinch region where $|\xi|$ is small we find good agreement between both the similarity solution and the full numerical solution.  Regarding the post pinch-off behaviour, Fig.~\ref{fig:ChannelPinchOff}$(b)$ shows a $\log$-$\log$ plot of the distance between point where pinch-off occurs and the tip of the recoiling bubble as a function of time after pinch-off, $t-t_0$. Again, we estimate a similarity exponent consistent with the theoretical prediction.

\subsection{Evolution of a Taylor-Saffman-like bubble} \label{sec:TaylorSaffmanBubble}

\begin{figure}
	\centering
	\includegraphics[width=0.80\linewidth]{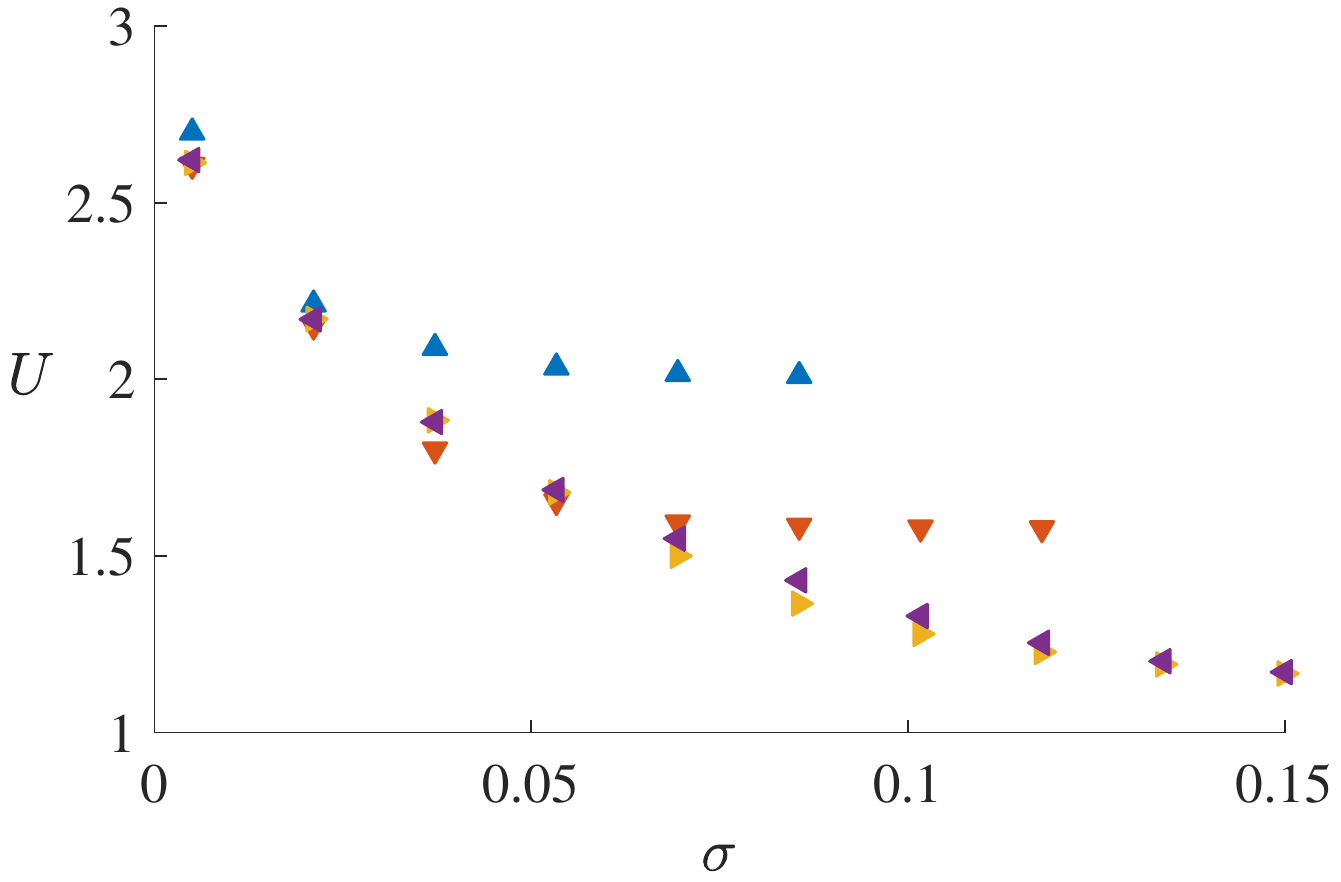}		
	\caption{The speed of an axially symmetric Taylor-Saffman-like bubble, $U$, propagating through a cylindrical tube as a function of the surface tension parameter $\sigma$. The initial condition is a spheroid with semi-axis given by $\rho = 0.4$ and $z = 0.2$ ($\blacktriangle$ blue), 0.4 ($\blacktriangledown$ red), 0.8 ($\blacktriangleleft$ yellow), and 1 ($\blacktriangleright$ purple). The bubble is evolved until both its shape and speed appear constant, at which time $U$ is estimated.}
	\label{fig:BubbleSpeed}
\end{figure}

In Sec.~\ref{sec:SaffmanTaylorFinger2} we discussed how an interface may break up to produce a bubble of finite volume that transverses through a cylindrical tube.  When such a bubble propagates along the channel and eventually settles into a steady motion with constant shape and speed $U$, it can be thought of as an axially symmetric version of a Taylor-Saffman bubble \cite{Taylor1959}. By choosing the initial condition of the bubble to be a prolate spheroid with varying aspect ratios, we are able to plot the steady-state speed $U$ versus surface tension $\sigma$ for a fixed volume of these Taylor-Saffman-like bubbles, as in Fig.~\ref{fig:BubbleSpeed}.  These plots show the same qualitative behaviour as the well-studied two-dimensional case \cite{Tanveer1986,Tanveer1987a}.
Namely, for small bubbles ($\blacktriangle$ and $\blacktriangledown$), the steady state shape of the bubble will tend to a sphere as $\sigma$ increases, whose speed is dependent upon the radius of this sphere. For larger bubbles whose size is too large to allow a spherical shape in the channel ($\blacktriangleleft$ and $\blacktriangleright$), their speed appears to be approaching unity as $\sigma$ increases, although it is not clear whether there is a maximum value of $\sigma$ for which solutions on this branch exist.  A difference between our axially symmetric bubbles and the two-dimensional Hele-Shaw problem is that in our case, each solution branch approaches $U=3^-$ in the limit $\sigma\rightarrow 0$, while in the two-dimensional case, the branches select $U=2^-$ in this limit.

\begin{figure}
	\centering
	\includegraphics[width=0.80\linewidth]{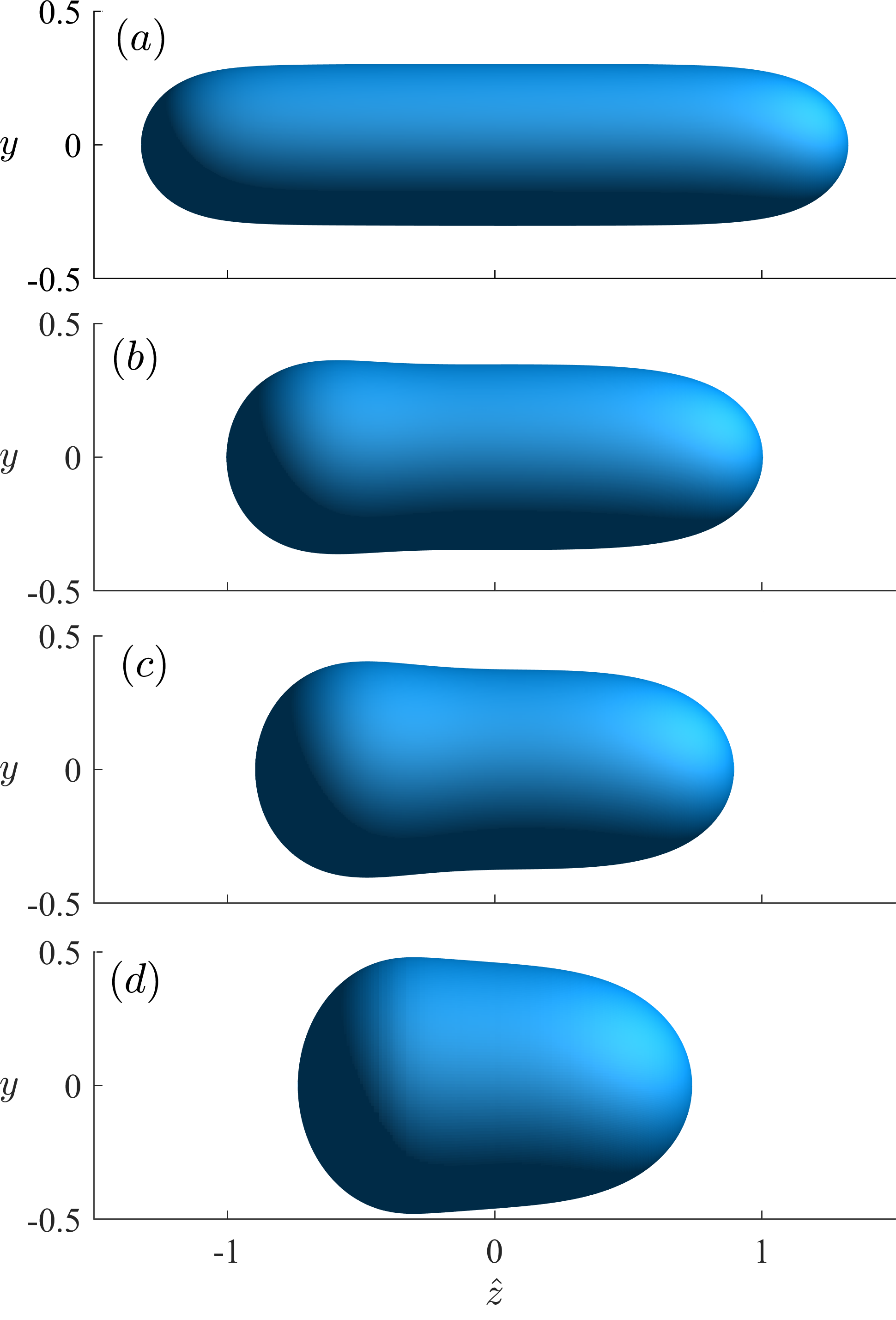}	
	\caption{Numerical simulation of bubble travelling through cylindrical tube with $\sigma$ $(a)$ $5 \times 10^{-4}$, $(b)$ $3 \times 10^{-3}$, $(c)$ $7 \times 10^{-3}$, and $(d)$ $1.5 \times 10^{-2}$. The initial condition of these simulations is a prolate spheroid of major and minor axes 1 and 0.4. The solutions are shown at a time sufficiently large such that both the speed and shape of the bubble are constant. The $\hat{z}$ coordinate is $z$ scaled such that the bubble is centred at the origin.}
	\label{fig:3dbubble}
\end{figure}

As just mentioned, while small steady bubbles will tend to spheres for large surface tension, the limiting shape of large steady bubbles as surface tension is increased is less obvious. We compare the numerical solution computed using different values of the surface tension parameter (at a sufficiently large time such that the bubble's speed and shape are essentially constant) in Fig.~\ref{fig:3dbubble}. For low surface tension (Fig.~\ref{fig:3dbubble}$(a)$), we find that the bubble tends to a long thin shape. As surface tension increases (Fig.~\ref{fig:3dbubble}$(b)$), we find that the sides of the bubble are no longer parallel but instead the bubble's trailing end is wider. For large values of surface tension (Fig.~\ref{fig:3dbubble}$(c)$ and $(d)$), we find the length of the bubble decreases and the width of the bubble is nearly as large as the radius of the cylinder.

The shape of the bubble in Fig.~\ref{fig:3dbubble}$(b)$ is similar to those of Taylor bubbles in cylindrical tubes.  This problem, where a large bubble is moving steadily in a tube otherwise filled with viscous fluid (in contrast to our problem, where the tube is filled with a porous medium saturated with viscous fluid), has an extensive history going back to Bretherton~\cite{Bretherton61} and Taylor~\cite{Taylor61} and is still attracting plenty of interest (see \cite{Balestra2018,Magnini19}, for example).  Our brief study in this section suggests that it is worth investigating further the links between our Taylor-Saffman-like bubbles and the well-studied Taylor bubbles.

\section{Discussion}\label{sec:discussion}

In this paper, we conducted a numerical and theoretical study investigating the behaviour of solutions to a one-phase Darcy model describing the evolution of a inviscid bubble within a fully saturated homogeneous porous medium. The model is a moving boundary problem which is an analogue of that for the one-phase Hele-Shaw problem, including a dynamic condition (\ref{eq:dynamic}) with pressure across the interface related to the curvature $\kappa$ via the surface tension.  This type of model is analysed and discussed in \cite{Ambrose2012,Ambrose2013,Brandao2018,Chuoke1959,Brener1993,Ceniceros2000,Dias2013,Levine1992,Cieplak1990,Martys1991,Rangel2009,Vondenhoff2009}, for example.  An important difference between our axially symmetric model and the Hele-Shaw version is that $\kappa$ represents the mean curvature of the surface while in the Hele-Shaw model it represents curvature in the plane.  This difference drives new behaviour not witnessed in the Hele-Shaw problem, which is worth recording.

A key focus of our paper is to study how a bubble undergoes a pinch-off singularity.  In Sec.~\ref{sec:NumericalScheme}, we summarised a flexible numerical scheme based on the level set method that is able to accurately solve \eqref{eq:Porous1}-\eqref{eq:Porous4} both before and after a pinch-off singularity has developed.  Using this scheme, we have shown that the self-similar solution derived by applying an integral equation (motivated by~\cite{Eggers2007}) matches well with full numerical solutions to \eqref{eq:Porous1}-\eqref{eq:Porous4}.  This was done for both contracting bubbles in radial geometry (Sec.~\ref{sec:Radial}) and translating bubbles in cylindrical channel geometry (Sec.~\ref{sec:propagating}).

\begin{figure}[h]
	\centering
	\includegraphics[width=0.80\linewidth]{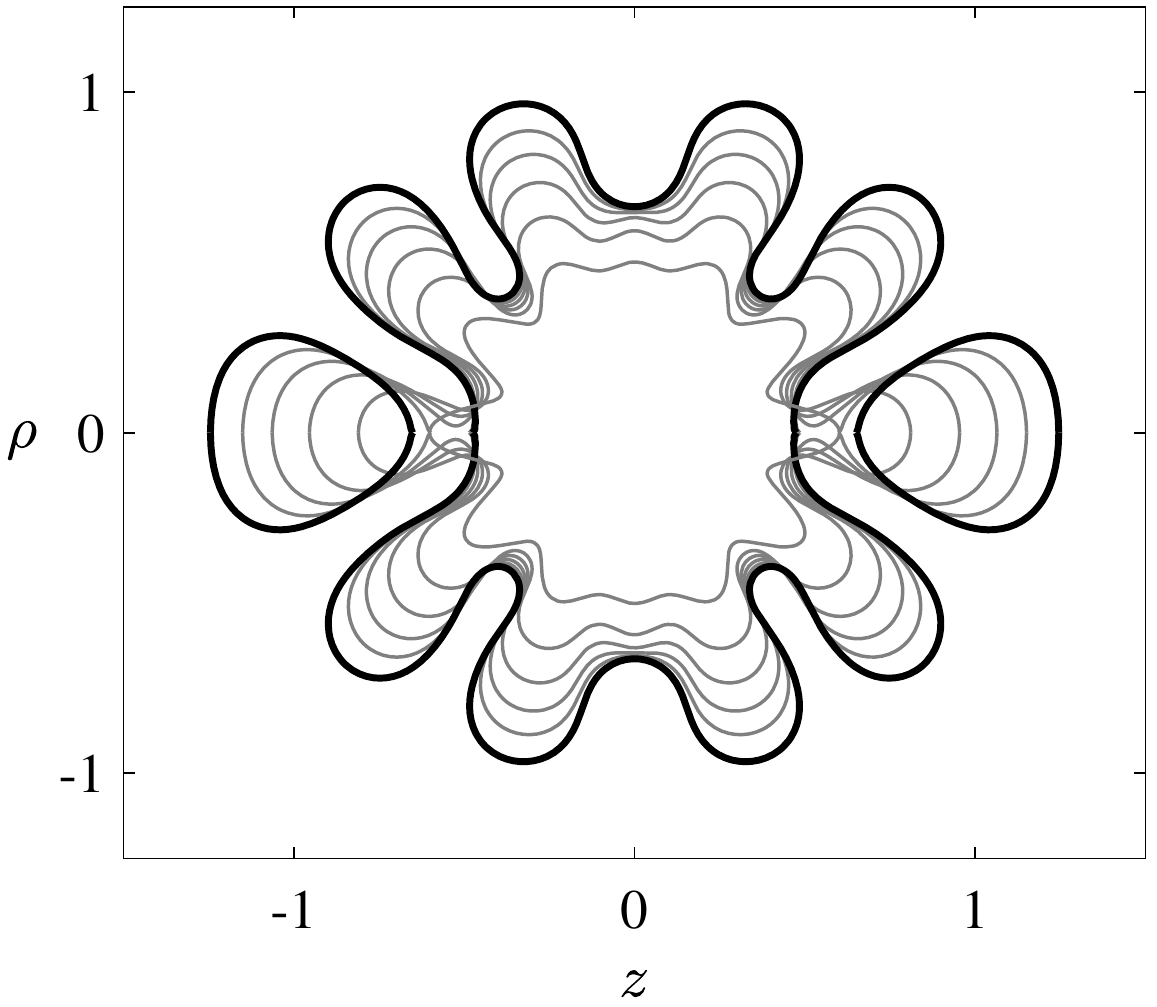}
	\caption{Numerical solution to \eqref{eq:Porous1}-\eqref{eq:Porous4} for an expanding bubble with $\sigma = 2.5 \times 10^{-3}$.  The interface of the bubble is unstable as it expands and a type of axially symmetric viscous fingering develops. When a single finger is aligned with the $z$-axis, it can pinch-off due to the curvature acting in the cylindrical direction.}
	\label{fig:viscousfingerpinchoff}
\end{figure}

It is worth mentioning the problem of an expanding bubble due to injecting an inviscid fluid at a point, as illustrated by Fig.~\ref{fig:Geometry}$(d)$.  This is a very well studied geometry in the two-dimensional Hele-Shaw case, as it acts as a prototype problem for viscous fingering and interfacial pattern formation \cite{Ben1990,Kessler1988,Langer1980}.  In three dimensions, the interfacial dynamics are more complicated due to the extra degree of freedom \cite{Brandao2018,Dias2013} and the possibility of fingers pinching off the main bubble in the same manner as we have been studying in this paper.  We illustrate this point via an example shown in Fig.~\ref{fig:viscousfingerpinchoff}.  Here we solve (\ref{eq:Porous1})-(\ref{eq:Porous4}) with the positive sign in (\ref{eq:Porous4}) in order to simulate a bubble expanding.  For the particular initial condition used in Fig.~\ref{fig:viscousfingerpinchoff}, we see a finger propagate along the positive $z$-axis which eventually breaks off the main bubble.  While we have not included the details here, the accompanying singularity in curvature is of the same form, both before and after pinch-off, as that considered in Sections \ref{sec:Radial} and \ref{sec:propagating}.  Of course, this example is partly contrived as it involves an axially symmetric geometry that does not illustrate the full three-dimensional nature of viscous fingering.  It does, however, demonstrate how singularities in curvature develop for expanding bubbles, a feature which is not present in two dimensions.

While our main physical motivation has been to study the evolution of a bubble in a porous medium, as noted in Sec.~\ref{sec:StefanProblem}, \eqref{eq:Porous1}-\eqref{eq:Porous4} can also be interpreted as describing the melting/freezing of a crystal dendrite. The pinch-off of crystals has been experimentally studied in \cite{Ishiguro2007}, where it was shown that the neck of the crystal both approaches and recoils from pinch-off with the scaling exponent $\alpha = 1/3$. To make analytical progress, the authors argued that the solid-melt interface can be approximated via mean curvature flow, $v_n = -\kappa$, which was shown to have a similarity exponent of $\alpha = 1/2$. The authors offer the explanation for this discrepancy that the mean curvature flow is slow to converge to self-similar behaviour, and the experimental results were not accurate enough to reflect this. It is interesting that our model \eqref{eq:Porous1}-\eqref{eq:Porous4}, which is the large Stefan number version of (\ref{eq:HeatEqn}) with \eqref{eq:Porous2}-\eqref{eq:Porous4}, gives the same similarity scaling.

Apart from studying pinch-off singularities, our other key focus has been on the time-dependent development of axially symmetric analogues of Saffman-Taylor fingers and Taylor-Saffman bubbles in a cylindrical channel, highlighting the differences and similarities between the axially symmetric and two-dimensional problems.  One noteworthy property of the finger problem is that families of steadily propagating axially symmetric fingers merge for finite values of surface tension in a similar manner to that observed in the Hele-Shaw wedge problem \cite{Amar1991}.  Our numerical results indicate that the speed of bubbles (of any volume) will tend towards $3^{-}$ in the limit that $\sigma \to 0$, which differs from Taylor-Saffman bubbles in a two-dimensional Hele-Shaw cell whose speed is known to tend towards $2^{-}$ \cite{GreenLustriMcCue,Lustri2018selection,Tanveer1986,Tanveer1987a,Tanveer1989}.   More accurate calculations of the data in Fig.~\ref{fig:BubbleSpeed} and shapes of steady-state axially symmetric Taylor-Saffman-like bubbles could be obtained by moving to the reference frame of the steadily moving bubbles and applying a boundary integral method.  The accompanying selection problem for $U=3$ would provide an interesting challenge requiring techniques in exponential asymptotics, especially given the lack of exact solution for the leading order problem.

\vspace{3ex}
\section*{Acknowledgements} SWM and LCM acknowledge the support of the Australian Research Council Discovery Project DP140100933.  We are grateful to the anonymous referees for their helpful feedback.

\bibliography{BibFile}

\appendix

\section{Prescribed rate of volume decrease} \label{sec:ROCvolume}

We now show that the form of the far-field boundary condition \eqref{eq:Porous4} results in the rate of change of volume of a bubble evolving according to \eqref{eq:Porous1}-\eqref{eq:Porous4} to be $\pm 4 \pi$.
%If a change in topology occurs, the total rate of change of volume will remain $\pm 4 \pi$, however each of the satellite bubbles formed after pinch-off may expand or contract at a different rate.
%For the two-dimensional analogue to \eqref{eq:Porous1}-\eqref{eq:Porous4}, where the farfield condition is $\partial u / \partial r \sim -Q / r$ as $r \to \infty$, the rate of change of area of the bubble can be shown to be $-2 \pi Q$. Here we show that the rate of change of volume of the bubble governed by \eqref{eq:Porous1}-\eqref{eq:Porous4} is $\dot{V} = -4 \pi Q$.
The rate of change of volume of a hypersurface whose boundary moves with normal velocity $v_n$ is
\begin{align}
\frac{\textrm{d} V}{\textrm{d} t} = \int_{\partial \Omega} v_n  \,\textrm{d} S,
\end{align}
which, using  the kinematic boundary condition \eqref{eq:Porous2}, gives
\begin{align} \label{eq:CoV1}
\frac{\textrm{d} V}{\textrm{d} t} = \int_{\partial \Omega} \grad \phi \cdot \bmth{n} \,\textrm{d} S.
\end{align}
Supposing that the viscous fluid occupies the region $\bmth{x} \in B \backslash \Omega(t)$, where $B$ is a sphere of radius $R$, then
\begin{align} \label{eq:CoV2}
0 = \int_{B \setminus \Omega} \nabla^2 \phi \, \text{d} V = \int_{\partial B} \grad \phi \cdot \bmth{n} \, \textrm{d} S - \int_{\partial \Omega} \grad \phi \cdot \bmth{n} \, \textrm{d}S,
\end{align}
which comes about via the divergence theorem.  Setting $\nabla^2\phi=0$ and substituting \eqref{eq:CoV1} into \eqref{eq:CoV2} gives
\begin{align}
\frac{\textrm{d}V}{\textrm{d}t} = \int_{\partial B}^{} \nabla \phi \cdot \bmth{n} \, \textrm{d}S = \int_{\partial B} \pm \frac{1}{r^2} \, \textrm{d}S = \pm 4 \pi.
\end{align}
We note that if a bubble were to undergo a change in topology, the  rate of change of the total volume will remain $\pm 4 \pi$. However, each individual bubble may not expand or contract at a constant rate. We further investigate the evolution of bubbles after pinch-off has occurred in Sec.~\ref{sec:NumericalScheme}.

\end{document}